\documentclass[twocolumn,english,showpacs,floatfix,amsmath,amssymb,prb,superscriptaddress]{revtex4-1}
\usepackage[dvips]{graphicx}
\usepackage{epsf}
\usepackage{psfrag}
\usepackage[colorlinks=true,citecolor=blue,linkcolor=red,linktocpage=true,pagebackref=false]{hyperref}
\usepackage{color,soul}

\newcommand{\be}{\begin{equation}}
\newcommand{\ee}{  \end{equation}}
\newcommand{\ba}{\begin{eqnarray}}
\newcommand{\ea}{  \end{eqnarray}}

\begin{document}

\title{Transport phenomena in helical edge states interferometers. A Green's function approach.}

\author{Bruno Rizzo} 
\affiliation{Departamento de Fisica and IFIBA, Facultad de Ciencias Exactas y Naturales, Universidad de Buenos Aires, Pab.\ I, Ciudad Universitaria, 1428 Buenos Aires, Argentina }

 \author{Liliana Arrachea} 
\affiliation{Departamento de Fisica and IFIBA, Facultad de Ciencias Exactas y Naturales, Universidad de Buenos Aires, Pab.\ I, Ciudad Universitaria, 1428 Buenos Aires, Argentina }
 
 \author{Michael Moskalets}
\affiliation{Department of Metal and Semiconductor Physics, NTU ''Kharkiv Polytechnic Institute'', 61002 Kharkiv, Ukraine}

\date{\today}
\begin{abstract}
We analyze the current and the shot-noise of an electron interferometer made of the helical edge states of a two-dimensional topological insulator
within the framework of non-equilibrium Green's functions formalism. We study in detail setups with a single and with two quantum point contacts inducing scattering between
the different edge states. We consider processes preserving the spin as well as the   
 effect of spin-flip scattering. In the case of a  single quantum point contact, a simple test based on the shot-noise measurement is proposed to quantify the strength of the spin-flip scattering.     
 In the case of two single point contacts with the additional ingredient of gate voltages applied within a finite-size region at the top and bottom edges of the sample, we identify two type of interference 
 processes in the behavior of the currents and the noise. One of such processes is analogous to that taking place in a Fabry-P\'erot interferometer, while the second one corresponds to a
 configuration similar to a Mach-Zehnder interferometer. In the helical interferometer these two processes compete.
\end{abstract}
\pacs{73.23.-b, 73.50.Td, 73.22.Dj}
\maketitle

\section{Introduction}

Quantum Spin Hall (QSH) insulators \cite{Murakami:2004th,Bernevig:2006ij,Hasan:2010ku} support helical states at their edges (HES). \cite{Kane:2005hl,Wu:2006ds,Xu:2006da}
These are Kramers' pairs of counter-propagating electron states with opposite spin and, therefore, they are topologically protected \cite{Kane:2005gb} against disorder in the absence of time-reversal symmetry breaking factors such as a magnetic field \cite{Maciejko:2010cl} or magnetic impurities \cite{Maciejko:2009kw,Tanaka:2011kk,Hattori:2011hj}. 
Recent experiments performed on mercury telluride quantum wells in the QSH regime clearly demonstrated a dissipationless  charge transport through the helical edge states. 
\cite{Konig:2007hs,Konig:2008bz,Roth:2009bg} 
Due to the fact that they appear in the form of Kramer's pairs, the transport properties of these states is  strikingly different \cite{Buttiker:2009bg} from the transport properties of other topologically protected states, like  the chiral edge states  of a system \cite{Halperin:1982tb,Buttiker:1988vk} in the Quantum Hall  state\cite{Klitzing:1980kw}.

\begin{figure}[t]
 \centering
 \includegraphics[width=8cm]{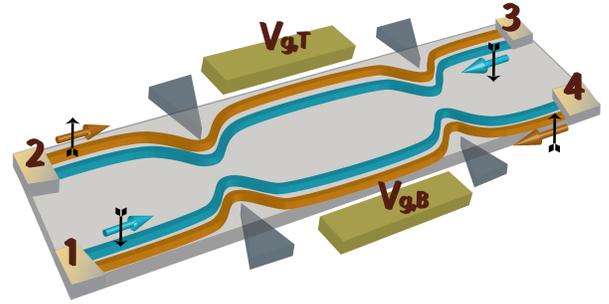}
 \caption{(Color online) Sketch of the topological insulator with two constrictions generating $M=2$ quantum point contacts at the positions $x_1,x_2$. The filling of the different edge states is globally 
 modified by recourse to four bias voltages $V_1, \ldots, V_4$. 
 In addition, two gate voltages applied to the top $V_{g,T}$ and bottom $V_{g,B}$ boundaries of the sample may locally modify the filling of the edge states within a finite region.}
\label{fig1}
\end{figure}

As a consequence of their helical nature, the edge states of the QSH insulators allow for the electrical control of spin currents. 
This property makes them promising for spintronic devices for quantum information processing. \cite{Prinz:1998bv,Wolf:2001fu}
A prominent feature of the HES is a strong correlation between the propagation direction and the spin of an electron, where neither spin nor the direction of movement are preserved separately. 
To account for this effect in transport through the HES we develop the corresponding general formalism in the framework of non-equilibrium Green's functions, which is adequate  to describe transport away from the linear response regime.  
As an example, we apply this formalism to analyze the correlation properties of currents flowing through two simple non-trivial helical circuits, corresponding to an interferometer comprising two branches with one and two quantum point contacts (QPC), Fig.~\ref{fig1}. 
The scattering at the QPCs connecting two helical states was discussed in detail in Refs.~\onlinecite{Hou:2009if,Strom:2009hj,Teo:2009bf,smith,orth}, while the effect of a wide tunneling contact was also recently analyzed. \cite{Dolcetto:2012bx,dolcini2013}

Quite generally the (local) time-reversal invariance only allows for the scattering between two helical states while the (back-)scattering within the same helical state (the same Kramers' pair) is forbidden even in the presence of a constriction. 
Therefore, the scattering at the QPC is characterized by two parameters, one describing  a spin-preserving scattering and one describing a spin-flip scattering between two Kramers' pairs. 
This type of a spin-flip scattering does not contradict to the (local) time-reversal invariance and it can take place even if more than one helical state exists at the same edge (in our case in the vicinity of the constriction). \cite{Xu:2006da,Konig:2008bz}  
The possibility of spin-flip processes makes an helical interferometer different from two independent copies of a chiral electronic interferometer like those built in the quantum Hall regime.
Interestingly, the helical interferometer shares \cite{Dolcini:2011dp}  properties of both the Mach-Zehnder  \cite{Ji:2003ck} and the Fabry-P\'erot interferometers  \cite{vanWees:1989bs,Camino:2007io}. 
Another feature, which makes a helical interferometer a non-trivial circuit is the possibility  of generating  an effective back-scattering processes within the same Kramers' pair. 
For the latter mechanism to take place the global time-reversal invariance has to be broken, for instance, by applying a weak magnetic field (which does not break the local time-reversal invariance).\cite{Delplace:2012ia} 
The realization of these processes rely on the presence of inter-helical-states spin-flip tunneling. 
Therefore, it  is crucial to know whether such processes are actually present  or no in real setups.  
In this paper we propose a  simple test based on the noise measurement, which enables to check the presence or absence of spin-flip tunneling between helical states at the QPC.

The paper is organized as follows. 
In Sec.~\ref{theor}  we describe the model of the helical interferometer comprising four metallic contacts as electron sources and sinks, helical edge states as electron waveguides, and some number $M$ of QPCs as wave splitters. Here we also present general equations for the currents and the current correlation functions. 
In Sec.~\ref{gfa} we calculate the transmission coefficients for the interferometer under study by using the Green's function approach. 
In Sec.~\ref{res} we present and discuss results for the current and the current correlation functions in circuits comprising one or two QPCs. 
We conclude in Sec.~\ref{concl}.

\section{Theoretical description}
\label{theor}
\subsection{Model}
We consider 
the setup addressed in
Refs. \onlinecite{Dolcini:2011dp,Citro:2011fw,romeo,Ferraro:2013du}. We model the TI as a ribbon with infinite length. 
Each longitudinal edge of the TI hosts a Kramers pair of  edge states, which is described by a Hamiltonian  of one-dimensional (1D) free electrons
with a definite helicity. The sample is assumed to have a number $M$ of constrictions that define QPCs through which electrons perform inter-pair
tunneling processes. Each constriction has two tunneling channels. One of them is spin preserving and the other one involves a spin-flip.

The full Hamiltonian reads 
\be
H=H_0+H_t+H_{f}+H_{g}.
\ee
The Hamiltonian for the HESs is 
\ba
H_0 &=&-i \hbar v_F \sum_{\sigma=\uparrow,\downarrow}\int dx [ : \Psi_{R,\sigma}^{\dagger}(x) \partial_x \Psi_{R,\sigma}(x) :\nonumber \\
& & -  : \Psi_{L,\overline{\sigma}}^{\dagger}(x) \partial_x \Psi_{L,\overline{\sigma}}(x)  : ],
\ea
where $\overline{\sigma}$ stands for the spin opposite to $\sigma$. 
We have assumed that a Kramers pair of right-moving($R$)  with $\uparrow$ spin electron states and left-moving ($L$) with $\downarrow$ spin ones
lie along the top edge of the sample, while another pair  $L, \uparrow$ and $R,\downarrow$ lie on the bottom, as shown in Fig \ref{fig1}. The Fermi velocity $v_F$ is assumed to be the same
for the two pairs and $:O:$ denotes normal ordering of the operator $O$.

The two types of tunneling terms represented by distinct quantum point contacts (QPC) located at $x_j, \; j=1,\ldots, M$, are 
\ba
H_{p} & = & \sum_{\sigma=\uparrow,\downarrow}\int dx \Gamma_{p}(x) [ \Psi_{R,\sigma}^{\dagger}(x) \Psi_{L,\sigma}(x) + 
 H. c. ],
\ea
for the spin-preserving process and
\be
H_{f} =  \sum_{\alpha {= L,R}}s_{\alpha}\int dx \Gamma_{f}(x) [\Psi_{\alpha,\uparrow}^{\dagger}(x)\Psi_{\alpha,\downarrow}(x)+ H. c.],
\ee
with $s_{R,(L)}=+ (-)$, for the case of spin-flip tunneling. 
We assume, for simplicity, that the corresponding amplitudes are the same for all the QPCs. Hence 
\be
 \Gamma_{p(f)}(x)= 2 \hbar v_F  \sum_{j=1}^M \gamma_{p(f)} \delta(x-x_{j}).
 \label{gamma}
\ee
The $H_g$ term refers to gate voltages applied at the top $V_{g,T}$ and bottom $V_{g,B}$ edges of the sample
 allowing for the manipulation of the filling of the helical channels 
within a finite  region  between the longitudinal coordinates $x_1$ and $x_M$,
\ba
H_{g} & = &  \int_{x_1}^{x_M} dx [ e V_{g,T} \left( \rho_{R,\uparrow}(x)+\rho_{L,\downarrow}(x)\right) \nonumber \\
& & + 
e V_{g,B} \left(\rho_{R,\downarrow}(x)+\rho_{L,\uparrow}(x)\right) ],
\ea
where $\hat{\rho}_{\alpha,\sigma}=:\Psi_{\alpha,\sigma}^{\dagger}(x)\Psi_{\alpha,\sigma}(x):$ is the local density operator of the species $\alpha, \sigma$ at the coordinate $x$.

\subsection{Transport properties}

In the setup of Fig.~\ref{fig1}, the transport is induced by changing the population of the edge states  by applying voltages at the four metallic contacts (reservoirs) indicated in the corners.
The coupling between edge states and metallic reservoirs is in general a subtle issue and the transport properties strongly depend on the details of the contacts. \cite{cham-frad} Here, we assume that the contact is such that the edges are in equilibrium with the respective reservoir of departure. 
It is important to notice that this configuration enables the induction of currents even for vanishing tunneling couplings $\gamma_{p(f)}$. In that case, such currents are due to an imbalanced population of right and left movers and the carriers do not experience any kind of scattering process. For instance,
 a voltage difference $V_1-V_4$ induces a current  only in the terminals $1$ and $4$, which flows $1 \rightarrow 4$ or $4 \rightarrow 1$ for
 $V_1> V_4$ and $V_4 > V_1$, respectively. Similarly, a voltage difference $V_2- V_3$ induces a current in
the terminals  $2$ and $3$, which flows $2 \rightarrow 3$ or $3 \rightarrow 2$ for
 $V_2 > V_3$ and $V_3 > V_2$, respectively. Interestingly, due to the helical nature of the edge states, each current flowing through a given terminal has a net
 polarization. For finite values of $\gamma_{p(f)}$, a voltage difference between any two contacts
induces currents through the four terminals, which are the result of scattering and interference effects due to the inter-edge tunneling through the QPCs. In this section we derive expressions for the currents flowing through the different terminals.

\subsubsection{Current}
 
The current operator is defined from the conservation of the charge in a given terminal $l=1, \ldots, 4$
\be
\dot{N_l}= \partial_x \{ v_F \left[\hat{\rho}^l_+(x) \; - \; \hat{\rho}^l_-(x)\right] \},
\ee
where $x$ is a position within that terminal and $\hat{\rho}^l_{\pm}(x)$ is the density operator corresponding to the incoming (outgoing) fermionic species flowing at that position. 
Thus, the current operator for electrons flowing into the terminal $l$ reads
\be
\hat{I}^l(x)=  e v_F\left[\hat{\rho}^l_+ (x)\; - \; \hat{\rho}^l_-(x)\right].
\ee
The ensuing mean value  $I^l(x)=\langle \hat{I}^l(x) \rangle$ can be expressed as follows
\be \label{cur}
I^l(x)= 
-i e v_F \left[G^<_{l+,l+}(x, x;t,t) - G^<_{l-,l-}(x, x;t,t)\right].
\ee
We have introduced the lesser Green's functions
\be
G^<_{\alpha \sigma, \alpha^{\prime} \sigma^{\prime}}(x, x^{\prime};t,t^{\prime})=
i\langle \Psi_{\alpha^{\prime},\sigma^{\prime}}^{\dagger}(x^{\prime},t^{\prime})\Psi_{\alpha,\sigma}(x,t)\rangle,
\ee
where the couple of indices $\alpha, \sigma$, with $\alpha=L,R$ and $\sigma= \uparrow, \downarrow$, labels the incoming (outgoing) state $l+$ ($l-$) of the terminal $l$. Due to the conservation of the charge, the current $I^l(x)$ does not depend on $x$ inside a given terminal, i.e. for positions
$x$  at the right (left) of the last (first) QPC.

\subsubsection{Noise}

To characterize the fluctuations of the currents  away from their mean values at the  terminals $l,l^{\prime}$ we introduce the following correlation function \cite{Blanter:2000wi}
\ba \label{sn}
S_{l,l^{\prime}}(x,x^{\prime};t,t^{\prime}) &= &\frac{1 }{2 }\left[\langle\lbrace \hat{I}^l(x,t),\hat{I}^{l^{\prime}}(x^{\prime},t^{\prime})\rbrace
 \rangle  \right.\nonumber \\
 & & \left. - 2\langle \hat{I}^{l}(x,t) \rangle \langle \hat{I}^{l^{\prime}}(x^{\prime},t^{\prime}) \rangle\right].
 \ea
The spectral power of current fluctuations, the noise power, reads
\be
S_{l,l^{\prime}}(x,x^{\prime},\omega)=\int_{-\infty}^{+\infty}d\tau e^{i\frac{\omega}{\hbar}\tau} S_{l,l^{\prime}}(x,x^{\prime};t,t+\tau), 
\label{noise0}
\ee
where $\tau=t-t^{\prime}$. 
We will study the $\omega=0$ component at $T=0$, called the  shot-noise.

For non-interacting fermions, the mean values entering Eq. (\ref{sn}) can be simply decoupled by recourse to Wick's theorem.
The resulting expression can be cast in terms of Fourier transforms of the lesser Green's functions\cite{bruno1}
\be
G^<_{\alpha \sigma, \alpha^{\prime} \sigma^{\prime}}(x, x^{\prime};t,t^{\prime})= \int_{-\infty}^{+\infty} \frac{d\omega}{2 \pi}
e^{-i \frac{\omega}{\hbar} (t-t^{\prime})} G^<_{\alpha \sigma, \alpha^{\prime} \sigma^{\prime}}(x, x^{\prime},\omega)
\ee
as follows
\ba
\label{noise}
& & S_{l,l^{\prime}}(0)=e^2 v_F^2\hbar \int_{-\infty}^{+\infty}\frac{d\omega}{2 \pi} \sum_{\xi ,\xi^{\prime} =+,-} \xi \xi^{\prime}  \left[ 
G_{l\xi ,l^{\prime}\xi^{\prime}}^{<}(x,x^{\prime},\omega) \times \right. \nonumber\\ 
&& \left. G_{{l^{\prime}\xi^{\prime}},l\xi }^{>}(x^{\prime},x,\omega) +G_{{l^{\prime}}\xi^{\prime} ,l\xi  }^{<}(x^{\prime},x,\omega)G_{l\xi,{l^{\prime}}\xi^{\prime}  }^{>}(x,x^{\prime},\omega)
\right].
\ea

\section{Green's functions approach}
\label{gfa}
\subsection{Dyson's equations}
It was shown in the previous section that all the observables of interest can be expressed in terms of  lesser Green's functions.  
In order to calculate the latter we use the Schwinger-Keldysh-Kadanoff-Baym technique. We
 introduce the retarded Green's function,
\be \label{ret}
G_{\alpha \sigma,\beta \sigma^{\prime}}^r(x,x^{\prime};t,t^{\prime})=-i\theta(t-t^{\prime})\langle \lbrace \Psi_{\beta,\sigma^{\prime}}^{\dagger}(x^{\prime},t^{\prime}),\Psi_{\alpha,\sigma}(x,t)\rbrace\rangle.
\ee
Following the standard procedure\cite{jauho}, we derive the Dyson's equation for this Green's function from the equation of motion
\ba
& & -i\hbar\partial_{t^{\prime}}G_{\alpha \sigma,\beta \sigma^{\prime}}^r(x,x^{\prime};t,t^{\prime})=\delta(t-t^{\prime})\delta_{\alpha,\beta}\delta_{\sigma,\sigma^{\prime}} \nonumber \\
& & -i\theta(t-t^{\prime})\langle \lbrace [H,\Psi_{\beta,\sigma^{\prime}}^{\dagger}(x^{\prime},t^{\prime})],\Psi_{\alpha,\sigma}(x,t)\rbrace\rangle.\label{eqofmotion}
\ea
Since the Hamiltonian contains spin-preserving and spin-flipping terms, the Dyson's equation reduces to a set of linear equations for 
$G_{\alpha \sigma,\beta \sigma^{\prime}}^r(x,x^{\prime},\omega)$, where the latter function is the
Fourier transform of (\ref{ret}). The corresponding equations for the retarded as well as the lesser function are explicitly shown in Appendix \ref{apa}.

In what follows, we discuss some formal elaboration of  the Dyson's equations, which is useful to evaluate the currents and to set the relation to the scattering matrix approach.
Notice that in order to evaluate the current from Eq. (\ref{cur}) we just need the diagonal elements of the Green function in the indices
 $\lambda \equiv \alpha, \sigma$. We start by  focusing on the diagonal elements of the retarded Green's function ${\cal G}_{\lambda}^r(x,x^{\prime},\omega) \equiv G_{\lambda,\lambda}^r(x,x^{\prime},\omega)$.
 After some algebra on the Dyson's Equation
based on back-substituting Green's functions with off-diagonal indices  (see Appendix A)  we obtain the following equation 
\ba\label{dysonr}
{\cal G}_{\lambda}^r(x,x^{\prime},\omega) &=&
g^{0,r}_{\lambda} (x,x^{\prime},\omega)+ \sum_{j,j^{\prime}=1}^M {\cal G}_{\lambda}^r(x,x_j,\omega) \nonumber \\
& & \times \Sigma_{\lambda}^r(x_j, x_{j^{\prime}},\omega) g^{0,r}_{\lambda} (x_{j^{\prime}},x^{\prime},\omega),
\ea
We have introduced the ''self-energy'' describing the escape of the electrons at 
edge $\lambda$ to the other edges. It is also convenient to cast the terms associated to the arguments
$x_j, x_{j^{\prime}}, j=1, \ldots, M$ as elements of $M \times M$ matrices. In this matrix notation, the self-energy reads
\be \label{self}
\hat{ \Sigma}_{\lambda}^r(\omega)  =   \hat{\Sigma}_{\lambda, \lambda}^{0,r}(\omega)+
\hat{\Sigma}_{\lambda, \overline{\lambda} }^{0,r}(\omega) \hat{G}^{0,r}_{\overline{\lambda}} (\omega)
\hat{\Sigma}_{\overline{\lambda}, \lambda}^{0,r}(\omega),
\ee
where the functions entering the matrices $\hat{\Sigma}_{\lambda, \lambda^{\prime}}^{0,r}(\omega)$ have been defined in Appendix \ref{apa}.
The Green's function 
$G^{0,r}_{\lambda} (x_j, x_{j^{\prime}},\omega), j,j^{\prime}=1,\ldots,M$ entering (\ref{self}) can also be organized in a matrix form as follows 
\be
\hat{G}^{0,r}_{\lambda} (\omega)=\left[ [ \hat{g}^{0,r}_{\lambda} (\omega) ]^{-1} - \hat{\Sigma}_{\lambda, \lambda}^{0,r}(\omega) \right]^{-1}.
\ee
The corresponding lesser Green's functions ${\cal G}_{\lambda}^<(x,x^{\prime},\omega)$ can be calculated by using Langreth rules\cite{jauho} in Eq. (\ref{dysonr}), as
discussed in
Appendix \ref{apa}. It is convenient to decompose this function as follows
\ba \label{gless}
{\cal G}^<_{\lambda}(x,x^{\prime},\omega) &= & {\cal G}^{<, {\rm eq}}_{\lambda}(x,x^{\prime},\omega) + \sum_{j,j^{\prime}}{\cal G}^{r}_{\lambda}(x,x_j,\omega) \nonumber \\
& & \times \Sigma^{<,{\rm neq}}_{\lambda}(x_j,x_{j^{\prime}},\omega) {\cal G}^{a}_{ \lambda}(x_{j^{\prime}},x^{\prime},\omega).
\ea
The first term is the {\em equilibrium} lesser Green's function corresponding to all the edges with the same chemical potential $\mu_{\lambda}$. The latter corresponds to the reservoir from where the 
electrons at the $\lambda$ edge  are injected. It reads
\ba \label{gleseq}
 {\cal G}^{<, {\rm eq}}_{\lambda}(x,x^{\prime},\omega) & =&   \left[ {\cal G}^a_{\lambda}(x,x^{\prime},\omega) - {\cal G}^r_{\lambda}(x,x^{\prime},\omega) \right] f_{\lambda}(\omega).
 \ea
 It will be useful to define a reference chemical
potential  $\mu^0_{\lambda}$ and a lesser function corresponding to all the edges at equilibrium with that chemical potential ${\cal G}^{<, {\rm eq}}_{\lambda}(x,x^{\prime},\omega) |_{\mu^0_{\lambda}} = \left[ {\cal G}^a_{\lambda}(x,x^{\prime},\omega) - {\cal G}^r_{\lambda}(x,x^{\prime},\omega) \right] f^0_{\lambda}(\omega)$.
Hence, the lesser function (\ref{gleseq}) can be, thus, rewritten as 
 \ba
 {\cal G}^{<, {\rm eq}}_{\lambda}(x,x^{\prime},\omega) & = &  {\cal G}^{<, {\rm eq}}_{\lambda}(x,x^{\prime},\omega) |_{\mu^0_{\lambda}} + \left[ f_{\lambda}(\omega)- f^0_{\lambda}(\omega) \right]  \nonumber \\
& &\times \left[ {\cal G}^a_{\lambda}(x,x^{\prime},\omega) - {\cal G}^r_{\lambda}(x,x^{\prime},\omega) \right] 
\ea
where $f_{\lambda}(\omega)$ and  $f^0_{\lambda}(\omega)$  are the Fermi functions corresponding to  the chemical potentials $\mu_{\lambda}$ and a reference chemical
potential  $\mu^0_{\lambda}$, respectively. 
The advanced Green's function is related to the retarded one through
$ {\cal G}^a_{\lambda}(x,x^{\prime},\omega) = [{\cal G}^r_{\lambda}(x^{\prime},x,\omega)]^*$.
The non-equilibrium part of the lesser self-energy
can be calculated from Eq. ( \ref{dysonl}) in Appendix \ref{apa} and reads
\begin{widetext}
\ba \label{sigmaleseq}
  \hat{\Sigma}^{<,{\rm neq}}_{\alpha,\sigma}(\omega)  &=&  i \left[f_{\overline{\alpha} \sigma}(\omega)-  f^0_{\alpha \sigma}(\omega)\right]
 \hat{\Gamma}^p_{\overline{\alpha} \sigma}(\omega) 
 +i \left[f_{\alpha \overline{\sigma}}(\omega)-  f^0_{\alpha \sigma}(\omega)\right]
 \hat{\Gamma}^f_{\alpha \overline{\sigma} }(\omega) 
 +
  i \left[f_{\overline{\alpha} \overline{\sigma}}(\omega)-  f^0_{\alpha \sigma}(\omega)\right] \hat{\Gamma}^{pf}_{\overline{\alpha} \overline{\sigma}}(\omega) ,
\ea
\end{widetext}
where, as before,  we are using the matrix notation to omit explicit reference to the coordinates of the contacts $x_j, x_{j^{\prime}}, j=1, \ldots, M$.
The ''hybridization'' matrices $\hat{\Gamma}_{\lambda}(\omega)$ are
\begin{widetext}
\ba \label{hybrid}
 \hat{\Gamma}^p_{\overline{\alpha} \sigma}(\omega)& = & 4\hbar^2 v_F^2
 \{  \gamma_p^2 \hat{\rho}^0_{\overline{\alpha} \sigma} (\omega)
 +2 \gamma_p \gamma_f  s_{\overline{\alpha}}  \hat{\rho}_{\overline{\alpha} \sigma}^{0}(\omega)
\mbox{Re} \left[\hat{\Lambda}^r_{\alpha \sigma, \overline{\alpha} \overline{\sigma}}(\omega) \right] +
\gamma_f^2  \hat{\Lambda}^r_{\alpha \sigma, \overline{\alpha} \overline{\sigma}}(\omega) 
\hat{\rho}^0_{\overline{\alpha} \sigma} (\omega)
\hat{\Lambda}^a_{\alpha \sigma, \overline{\alpha} \overline{\sigma}}(\omega) \}, \nonumber \\
 \hat{\Gamma}^f_{\alpha \overline{\sigma} }(\omega) & = & 4\hbar^2 v_F^2
 \{ \gamma_f^2 \hat{\rho}^0_{\alpha \overline{\sigma}}(\omega)+
 2  \gamma_p \gamma_f  s_{\alpha} \hat{\rho}^0_{\alpha \overline{\sigma}}(\omega)
\mbox{Re} \left[\hat{\Lambda}^r_{\alpha \sigma, \overline{\alpha} \overline{\sigma}}(\omega) \right] + \gamma_p^2 
\hat{\Lambda}^r_{\alpha \sigma, \overline{\alpha} \overline{\sigma}}(\omega)  \hat{\rho}_{\alpha \overline{\sigma}}^{0}(\omega)
\hat{\Lambda}^a_{\alpha \sigma, \overline{\alpha} \overline{\sigma}}(\omega) \},\nonumber \\
\hat{\Gamma}^{pf}_{\overline{\alpha} \overline{\sigma}}(\omega)& = &\hat{\Sigma}^{0,r}_{\alpha \sigma, \overline{\alpha} \overline{\sigma}}(\omega)
\left[\hat{\Lambda}^r_{\overline{\alpha} \overline{\sigma}, \overline{\alpha} \overline{\sigma}}(\omega)+ \hat{1} \right] \hat{\rho}^0_{\overline{\alpha} \overline{\sigma}}(\omega)  
\left[\hat{1}  +\hat{\Lambda}^a_{\overline{\alpha} \overline{\sigma}, \overline{\alpha} \overline{\sigma}}(\omega) 
 \right]\hat{\Sigma}^{0,r}_{\overline{\alpha} \overline{\sigma},\alpha \sigma}(\omega)
\ea
\end{widetext}
where $\hat{\rho}^0_{\alpha \sigma}(\omega)= i \left[\hat{g}^{0,r}_{\alpha \sigma}(\omega)-\hat{g}^{0,a}_{\alpha \sigma}(\omega) \right]$, and
\be
\hat{\Lambda}^r_{\alpha \sigma, \overline{\alpha} \overline{\sigma}}(\omega)=\hat{\Sigma}^{0,r}_{\alpha \sigma, \overline{\alpha} \overline{\sigma}}(\omega) \hat{G}^{0,r}_{\overline{\alpha} \overline{\sigma}}(\omega),
\ee
with $[\hat{\Lambda}^r_{\alpha \sigma, \overline{\alpha} \overline{\sigma}}(\omega)]^{\dagger}=\hat{\Lambda}^a_{ \overline{\alpha} \overline{\sigma},\alpha \sigma}(\omega)$.

\subsection{Transmission functions}
\label{gtf}
The charge current flowing through the terminal $l$ defined in Eq. (\ref{cur}) can be written in terms of the lesser function defined in (\ref{gless}) as 
\be
I^l(x)= -i e v_F \int_{-\infty}^{+\infty} \frac{ d \omega}{2 \pi} \left[ {\cal G}^{<}_{l+}(x,x,\omega)- {\cal G}^{<}_{l-}(x,x,\omega) \right].
\ee
In what follows we will eliminate the explicit reference to the coordinate $x$ in the current and we will simply label this quantity with the terminal index $l$. The retarded Green's functions depend on $x$ and we will take any value of this coordinate within the terminal under consideration.
After identifying the pair of indices $\alpha \sigma$ ($\overline{\alpha} \overline{\sigma}$) corresponding to the  incoming (outgoing) channel of the terminal
$l$ and defining the reference chemical potential as the one corresponding to the outgoing channel, i.e $\mu_{\alpha,\sigma}^0 \equiv \mu_{l-}$ we substitute the lesser function of Eq. (\ref{gless}) with the representation  defined in Eq. (\ref{gleseq})  and find
\begin{widetext}
\ba \label{curg}
I^l &= &I^{l}_0+ e v_F \sum_{j j^{\prime}} \int_{-\infty}^{+\infty} \frac{ d \omega}{2 \pi} 
{\cal G}_{\alpha \sigma}^r(x,x_j,\omega) \{ \Gamma^p_{\overline{\alpha} \sigma} (x_j,x_{j^{\prime}},\omega) 
\left[f_{\overline{\alpha} \sigma}(\omega) - f_{l-}(\omega) \right]+
\Gamma^f_{ \alpha \overline{\sigma} } (x_j,x_{j^{\prime}},\omega) 
\left[f_{\alpha \overline{\sigma}}(\omega) - f_{l-}(\omega) \right]
\nonumber \\
&  & - \left[ \Gamma^p_{\overline{\alpha} \sigma} (x_j,x_{j^{\prime}},\omega)  + \Gamma^f_{ \alpha \overline{\sigma} } (x_j,x_{j^{\prime}},\omega) \right]
\left[f_{\alpha \sigma}(\omega) - f_{l-}(\omega) \right]
\} {\cal G}^a_{\alpha \sigma}(x_{j^{\prime}},x,\omega)- e v_F \sum_{j j^{\prime}} \int_{-\infty}^{+\infty} \frac{ d \omega}{2 \pi} {\cal G}_{\overline{\alpha} \overline{\sigma}}^r(x,x_j,\omega) \nonumber \\
& & \times 
\{ \Gamma^p_{ \alpha \overline{\sigma}} (x_j,x_{j^{\prime}},\omega)  \left[f_{\alpha \overline{ \sigma}}(\omega) - f_{l-}(\omega) \right]+
\Gamma^f_{ \overline{\alpha} \sigma} (x_j,x_{j^{\prime}},\omega) 
  \left[f_{ \overline{\alpha} \sigma }(\omega) - f_{l-}(\omega) \right] \}
{\cal G}^a_{\overline{\alpha} \overline{\sigma}}(x_{j^{\prime}},x,\omega),
\ea
\end{widetext}
where the first term corresponds to the current {\em without tunneling} to the other edges
\ba
I^{l}_0 & = &-i e v_F \int_{-\infty}^{+\infty} \frac{ d \omega}{2 \pi} \left[ {\cal G}^{<,{\rm eq}}_{\alpha \sigma}(x,x,\omega)- {\cal G}^{<, {\rm eq}}_{\alpha \sigma}(x,x,\omega)|_{\mu_{l-} }\right] \nonumber \\
& = &e v_F \int_{-\infty}^{+\infty} \frac{ d \omega}{2 \pi}\left[f_{l+ }(\omega) - f_{l- }(\omega) \right]
\rho^0_{\alpha \sigma} (x,x,\omega).
\ea
At zero temperature, substituting (\ref{g0les}) into $\rho^0_{\alpha \sigma}(x,x,\omega)$ results in 
\be
I^{l}_0= \frac{e}{h}(\mu_{l + }- \mu_{l - }).
\ee
Due to the chiral nature of the electronic motion within the edges, the retarded Green's function ${\cal G}^r_{R(L), \sigma}(x,x_{j^{\prime}},\omega)$ for right (left)-moving electrons vanish for $x< x_1$ ($x> x_M$), where we assumed that the 
scattering region containing the point contacts extends within $[x_1, x_M]$. In addition, we must take into account that  the terminals placed at the right side of the scattering region($x>x_ M$) correspond to $l+= R \uparrow, \; R \downarrow$ for $l=3,4$, respectively, while those at the left side have $l+= L \uparrow, \; L \downarrow$ for
$l=1,2$, respectively.
Thus, the two last lines of Eq. (\ref{curg}) vanish.  For the same reason, the term  $\propto \hat{\Gamma}^{p f}$ of (\ref{sigmaleseq}) does not contribute and the expression of the current 
can be cast into the form of the Landauer-B\"{u}ttiker formula, see, e.g., Ref.~\onlinecite{Buttiker:1992vr},
\ba \label{curt}
I^l & = &   \frac{e}{\hbar}\sum_{l^{\prime}=1}^4 \int \frac{ d\omega}{2 \pi}  \{ T_{l+,l^{\prime}-}(\omega)  \left[ f_{l^{\prime}-}(\omega) - f_{l -}(\omega)\right] \},
\ea
where  $T_{l+,l^{\prime}-}(\omega),\; l^{\prime}-=1, \ldots, 4$ are {\em transmission functions} between the incoming channel $l+$ and the remaining four channels
through the tunneling contacts. For the model we are considering, these functions explicitly read
\ba \label{trans}
& & T_{l+, l^{\prime}-}(\omega)= \hbar v_F  \sum_{j j^{\prime}}{\cal G}^r_{\alpha \sigma }(x,x_j,\omega)  
\{ \delta_{l^{\prime},\overline{\alpha}\sigma} \Gamma^p_{\overline{\alpha}\sigma}(x_j, x_{j^{\prime}},\omega) \nonumber \\
& &\;\;\;\;\;\;\;\;+ \delta_{l^{\prime},\alpha \overline{\sigma}} \Gamma^f_{\alpha \overline{\sigma}}(x_j, x_{j^{\prime}},\omega) \}
  {\cal G}^a_{\alpha\sigma}(x_{j^{\prime}},x,\omega) +  \hbar v_F \delta_{l^{\prime}, \alpha \sigma}  \nonumber \\
& & \;\;\;\;\;\;\;\; \times \lbrace\rho^0_{\alpha \sigma} (x,x,\omega)-\sum_{j j^{\prime}}{\cal G}^r_{\alpha \sigma }(x,x_j,\omega)  
[ \Gamma^p_{\overline{\alpha}\sigma}(x_j, x_{j^{\prime}},\omega)\nonumber \\
& & \;\;\;\;\;\;\;\; +  \Gamma^f_{\alpha \overline{\sigma}}(x_j, x_{j^{\prime}},\omega) ]
  {\cal G}^a_{\alpha\sigma}(x_{j^{\prime}},x,\omega)\rbrace,
  \ea
  where, given $l+ \equiv \alpha, \sigma$, the first term corresponds to the transmission between the channel $l^{\prime}-= \overline{\alpha}, \sigma$ and the channel $l+ = \alpha, \sigma$, the second term corresponds
  to the transmission between $l^{\prime}-=\alpha,  \overline{\sigma}$ and the channel $l+\equiv \alpha, \sigma$, and the latter is the transmission function between $l^{\prime}-= \alpha, \sigma$ and $l+$.
  The functions $\Gamma(x_j, x_{j^{\prime}},\omega)$ are the matrix elements of the hybridization matrix
  $\hat{\Gamma}^{p}_{\overline{\alpha} \sigma}(\omega)$, for $l^{\prime}- \equiv \overline{\alpha} \sigma$ or the ones of the matrix
   $\hat{\Gamma}^{f}_{\alpha \overline{\sigma}}(\omega)$, for $l^{\prime}- \equiv \alpha \overline{\sigma}$. 

\subsection{Relation to the scattering matrix formalism} 
\label{rsm}
     
It is interesting to notice that the transmission functions defined in the previous section set  an explicit relation between the scattering matrix and the Green's function formalism.  In the case of transport through normal tunneling contacts between two reservoirs at different chemical potentials, Fisher and Lee's equation\cite{fisher-lee} provides such an   explicit relation. This equation has been generalized to harmonically time-dependent problems,\cite{arra-mos} but so far, it has not been analyzed in the context of transport  through edge states alone without tunnel coupling to the contacts. To establish such a relation for transport of helical edge states in bar geometry we proceed as follows.

  For simplicity, we start by considering a single QPC connecting the top and bottom edge states. In such case we can set the following identity between elements of the
  scattering matrix and the retarded diagonal (in the edge indices) elements of the Green's functions
  \be \label{s}
  {\cal S}_{l+,l^{\prime}-}= -i \sqrt{\hbar v_F} {\cal G}^r_{\alpha \sigma}(x,x_1,\omega) 
  \sqrt{\Gamma_{\alpha^{\prime}\sigma^{\prime}}(x_1,x_1,\omega)},
  \ee
  for $\alpha^{\prime} \sigma^{\prime} \neq \alpha \sigma$ and $ \alpha^{\prime} \sigma^{\prime} \neq \overline{\alpha} \overline{\sigma}$.   
We recall the terminal $l$ has an incoming edge state characterized by $\alpha \sigma$, which is denoted by $l+$, and an outgoing one characterized by $\overline{\alpha} \overline{\sigma}$, which would be denoted by $l-$, while $l^{\prime}$  injects an edge characterized by $\alpha^{\prime} \sigma^{\prime}$, which is denoted by $l^{\prime}-$.
Hence, the elements of the scattering matrix 
  defined in Eq. (\ref{s}) correspond to terminals $l, l^{\prime}$ such that $l$ is on the top (bottom) and $l^{\prime}$ is on the bottom (top) of the sample. We also
  recall that $x$ lies on the  $l$-th terminal, and $x_1$ is the position of the QPC where the tunneling contact between the edge connected to the $l$ reservoir
  and the edge connected to the $l^{\prime}$ one . We can easily identify this expression with the one proposed by Fisher and Lee. The hybridization function 
  $\Gamma_{l^{\prime}-}(x_1,x_1,\omega)$ denotes scattering processes due to the escape of the electrons from the edge injected into the reservoir $l$ and the edge leaving the reservoir $l^{\prime}$. 
  In the present case, the contact of the incoming edge state and the reservoir $l$ is assumed to be ideally ballistic, thus the hybridization matrix
  is just represented by $\hbar v_F$.   
  In addition, ${\cal G}^r_{\alpha \sigma}(x,x^{\prime},\omega) =0$ for $x^{\prime} >x $ within the $l$ terminal, which is a consequence of the absence
  of reflection of the helical edge state into the terminal of departure. In the present system, this is due to  time-reversal symmetry which dictates lack of scattering between the two states of a Kramer's pair. Thus,
  \be
  {\cal S}_{l+,l-}=0.
  \ee  
  In all the cases we define the transmission function
  as follows
  \be
  T_{l+,l^{\prime}-}= |{\cal S}_{l+,l^{\prime}-}|^2,
  \ee
  while the conservation of the charge impose
  \be
  \sum_{l^{\prime \prime}- } {\cal S}_{l+,l^{\prime \prime}-} {\cal S}_{l^{\prime}+,l^{\prime \prime}-}^{*}=\delta_{l,l^{\prime}}.
  \ee
  Thus,
  \be
  \sum_{l^{\prime}-}T_{l+,l^{\prime}-}=1. \label{cons}
  \ee
 
 In the case of $M$ QPC, the scattering between top and bottom edges becomes multichannel. In such case, Fisher and Lee's equation is more suitable represented as
 \ba
    T_{l+,l^{\prime}-} &=& \Gamma_{l+} \sum_{j,j^{\prime}=1}^M {\cal G}^r_{\alpha \sigma}(x,x_j,\omega) \Gamma_{\alpha^{\prime} \sigma^{\prime}}(x_j,x_{j^{\prime}},\omega) \nonumber \\
  & & \times  {\cal G}^a_{\alpha \sigma}(x_{j^{\prime}},\omega),    \ea
 for $\alpha^{\prime} \sigma^{\prime} \neq \alpha \sigma$ and $ \alpha^{\prime} \sigma^{\prime} \neq \overline{\alpha} \overline{\sigma}$. We consider a ballistic hybridization parameter  $\Gamma_{l +}= \hbar v_F$ associated to the ideal connection between the edge and
  the reservoir $l$ towards it travels, while the $M\times M$ hybridization matrix $\Gamma_{\alpha^{\prime} \sigma^{\prime}}(x_j,x_{j^{\prime}},\omega)$ represents
 the escape to the edge $\alpha^{\prime} \sigma^{\prime}$, which is injected from the reservoir $l^{\prime}$, through the $M$ QPCs. Due to the helical nature of the states
   $T_{l+,l-}=0$, while  due to the conservation of the charge  (\ref{cons}) 
   \be
   T_{l+,\overline{l}-}=1-\sum_{l^{\prime }- \neq l+} T_{l+,l^{\prime }-},
   \ee
   where $\overline{l}- \equiv l+$ denotes the reservoir injecting the $\alpha \sigma$ edge state that incomes into  the $l$ reservoir. The two reservoirs
   $l$ and $\overline{l}$ are on the same (top or bottom) part of the sample.  

\subsection{Currents for particular configuration of bias voltages}
In order to illustrate the use of the previous expressions we present more explicitly the expressions for the currents along the terminal $l=3$ for three different
configurations of  voltages.
\subsubsection{Charge-bias configuration}
This corresponds to 
\begin{eqnarray}
V_{1} &=& V_{2} =V \,,
\nonumber \\ 
\label{cbcv} \\
V_{3} &=& V_{4} =0 \,.
\nonumber
\end{eqnarray}
We, thus, assume $\mu_2=\mu_1= \mu+ eV$ and
$\mu_3=\mu_4=\mu$. 
Eq. (\ref{curt}) for the current through $l=3$ corresponds to $l+ \equiv R, \uparrow$ and $l- \equiv L, \downarrow$
\ba
I^3 &= & \frac{e}{\hbar}\int \frac{ d\omega}{2 \pi} \{T_{3,1}(\omega)\left[f_1(\omega)-f_3(\omega) \right]  
\nonumber \\
 & & +T_{3,2 }(\omega)\left[f_2(\omega)-f_3(\omega)\right]
  \},
  \label{cbc}
\ea
with 
\ba
T_{3,1}& &=  \hbar v_F \sum_{j j^{\prime}}{\cal G}^r_{R \uparrow }(x,x_j,\omega) \Gamma^p_{L,\uparrow} (x_j, x_{j^{\prime}},\omega)  {\cal G}^a_{R \uparrow }(x_{j^{\prime}},x,\omega),\nonumber \\
T_{3,2}& & = \hbar v_F \rho^0_{R \uparrow} (x,x,\omega)-\hbar v_F\sum_{j j^{\prime}}{\cal G}^r_{R \uparrow }(x,x_j,\omega) \times  \nonumber \\
& & \left[ \Gamma^p_{L\uparrow}(x_j, x_{j^{\prime}},\omega)+  \Gamma^f_{R \downarrow}(x_j, x_{j^{\prime}},\omega) \right]
  {\cal G}^a_{R\uparrow}(x_{j^{\prime}},x,\omega),
  \label{T312}
\ea
where we have simplified the notation and expressed $T_{3,l^{\prime}}, \; l^{\prime}=1,2$, instead of $T_{3+, l^{\prime}-}$ as in the previous section.
This configuration generates a net longitudinal flow of charge.

Another possibility is to generate net transverse charge flow,
\begin{eqnarray}
V_{1} &=& V_{4} =V \,,
\nonumber \\ 
\label{ntct} \\
V_{2} &=& V_{3} =0 \,.
\nonumber
\end{eqnarray}
This corresponds to $\mu_2= \mu_3= \mu$, and
 $\mu_1=\mu_4=\mu+eV$. In this case the current through $l=3$ reads
\ba
I^3 &= & \frac{e}{\hbar}\int \frac{ d\omega}{2 \pi} \{T_{3,1}(\omega)\left[f_1(\omega)-f_3(\omega) \right]  
\nonumber \\
 & & +T_{3,4 }(\omega)\left[f_4(\omega)-f_3(\omega)\right]  \},
\ea
with
\be
T_{3,4} =\hbar v_F \sum_{j j^{\prime}}{\cal G}^r_{R \uparrow }(x,x_j,\omega) \Gamma^f_{R,\downarrow} (x_j, x_{j^{\prime}},\omega)  {\cal G}^a_{R \uparrow }(x_{j^{\prime}},x,\omega). 
\label{T34}
\ee

\subsubsection{Spin-bias configuration}
Here we have 
\begin{eqnarray}
V_{1} &=& -V_{2} =V \,,
\nonumber \\ 
\label{ntcs} \\
V_{3} &=& V_{4} =0 \,.
\nonumber
\end{eqnarray}
We, thus, assume 
$\mu_1= \mu+ eV$, $\mu_2=\mu-eV$ 
and
$\mu_3=\mu_4=\mu$. In this case the current through $l=3$ 
is given in Eq.~(\ref{cbc}).

Another spin-bias configuration corresponds to 
\begin{eqnarray}
V_{2} &=& V_{4} =V \,,
\nonumber \\ 
\label{ntc} \\
V_{1} &=& V_{3} =0 \,,
\nonumber
\end{eqnarray}
in which case the current through $l=3$ reads
\ba
I^3 &= &\frac{e}{\hbar}\int \frac{ d\omega}{2 \pi} \{T_{3,2}(\omega)\left[ f_2(\omega)-f_3(\omega) \right]  
\nonumber \\
 & & +T_{3,4 }(\omega)\left[f_4(\omega)-f_3(\omega)\right]
  \}.
\ea

\section{Results}
\label{res}

We now show results for the behavior of the current and the noise power. We focus on temperature $T=0$ and $\mu=0$, and we analyze separately the case of a single QPC 
and two QPCs. We have verified that we recover the results presented in Ref. \onlinecite{Dolcini:2011dp} for the first type of charge and spin bias configurations
presented in the previous section. 

In what follows, we discuss the second type of spin-bias configurations shown in the previous section,
$V_2=V_4=V$ and $V_1=V_3=0$, Eq.~(\ref{ntc}), which has not been studied in previous works. \cite{Dolcini:2011dp,romeo}
In some cases, we extend our study to a more general configuration with $V_2\neq V_4$ and $V_1=V_3=0$.

\subsection{Single point contact}

In the case of a single QPC, the transport is not affected by the top and bottom gate voltages $V_{g, T}$ and $V_{g,B}$.  The transmission functions entering the current through the different terminals can be easily evaluated by substituting Eq. (\ref{gr1}) in Eq. (\ref{trans}). The final expressions are summarized for completeness in Appendix D.1.

\subsubsection{Currents}

In Fig.~\ref{fig2} we show the behavior of the currents through the terminals $l=3,4$ vs the spin-flipping tunneling amplitude $\gamma_f$ and a finite spin-preserving
tunneling $\gamma_p$ at the QPC. 

\begin{figure}
        \centering
        \includegraphics[width=8cm]{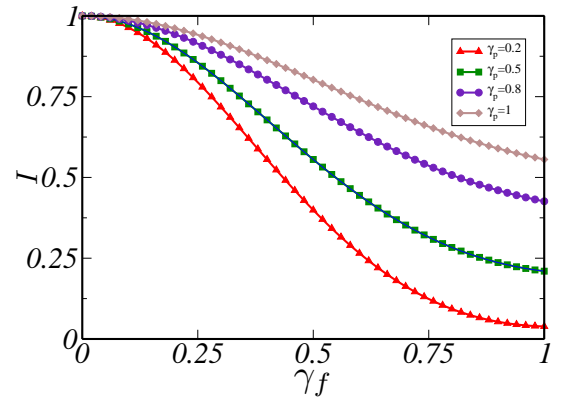}
 \caption{(Color online) 
Current $I_{1}=I_{3}=-I_2=-I_4$  in units of $e^{2} V /h$ versus spin-flipping term in a spin-bias configuration $V_1=V_3=0$ and $V_2=V_4=V$.
Different plots correspond to different values of the  spin-preserving tunneling $\gamma_{p}$. } 
\label{fig2}
\end{figure}

Notice that because of the symmetry of the setup $I_1=I_3$, $I_2=I_4$ and $I_3= -I_4$. For vanishing $\gamma_f$ and $\gamma_ p$ there is a net $\uparrow$  flow from the left to the right in the upper terminals, and another net $\uparrow$ flow in the lower terminals and a vanishing current through the QPC. 
Turning on just the spin preserving
tunneling ($\gamma_p \neq 0, \; \gamma_f=0$) for the present bias configuration does not change this picture. In fact, the contacts $2$ and $4$ inject an identical flow of $\uparrow$ 
electrons that travel in opposite directions and the spin preserving tunneling does not change the zero value of the net current flowing through the QPC due to the Pauli exclusion principle. This situation, however, changes with a finite  spin flip tunneling $\gamma_f \neq 0$. This tunneling process effectively enables the transmission of  electrons injected from the contact $2$  into the contact $4$ after flipping the spin
$\uparrow \rightarrow \downarrow$ at the QPC. Conversely, electrons injected with $\uparrow$ spin from $4$ perform a spin flip tunneling at the QPC and enter the contact 
$2$ with spin $\downarrow$.  The net result of these processes is a decreasing net current through all the terminals as well as through the QPC. In particular, 
for small $\gamma_p$ and $\gamma_f  \sim 1$, the QPC behaves as an approximately ballistic contact for each spin species, allowing for a perfect transfer of particles  accompanied by a spin-flip. For this bias
configuration, however, the top-to-bottom flow is identical to the bottom-to-top one and they cancel one another, resulting in a vanishing small net current through the QPC, as well as through the four terminals.

\subsubsection{Noise}

\begin{figure}
         \centering
        \includegraphics[width=8cm]{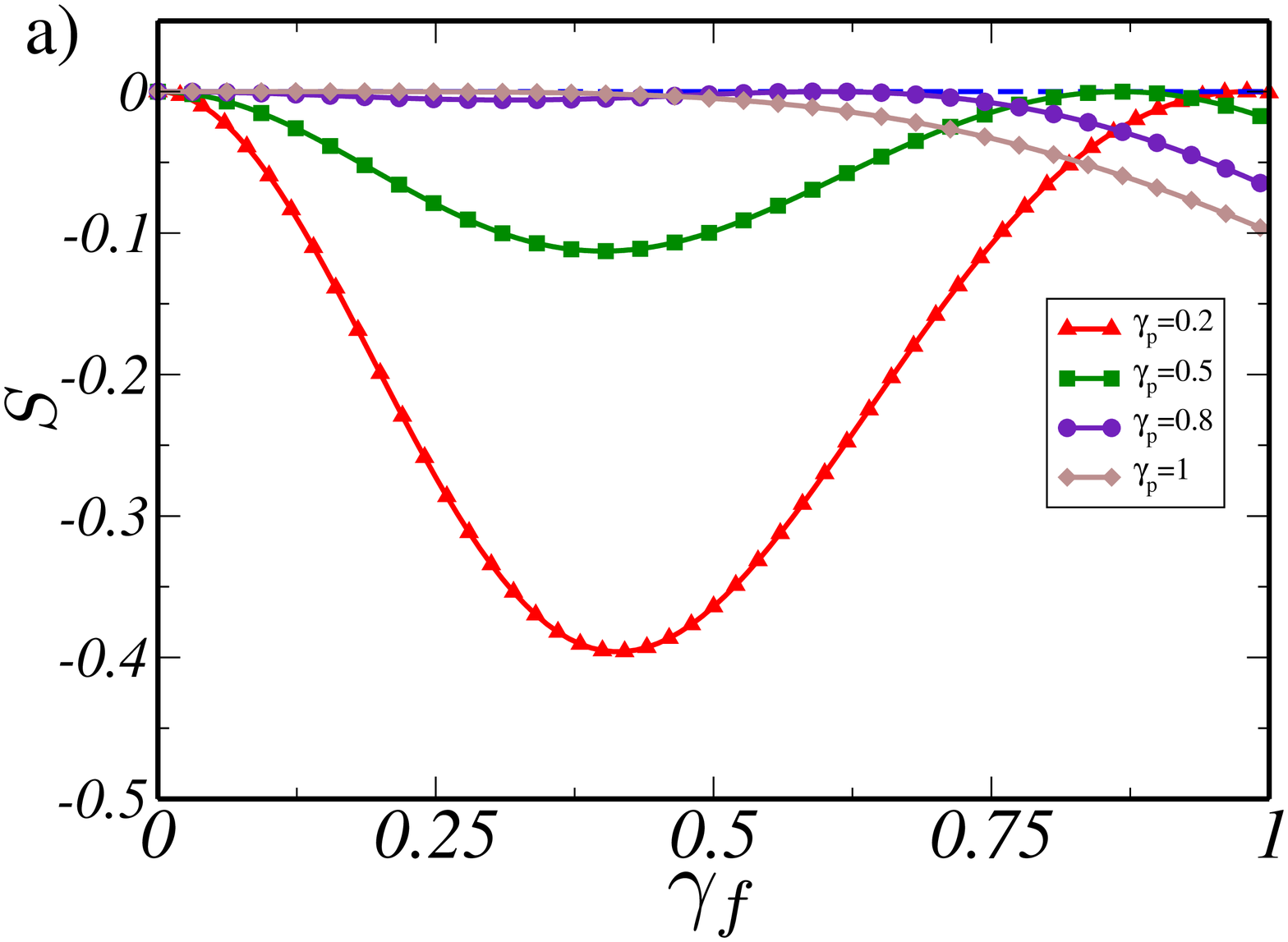}\vspace{0.7cm}\\
        \includegraphics[width=8cm]{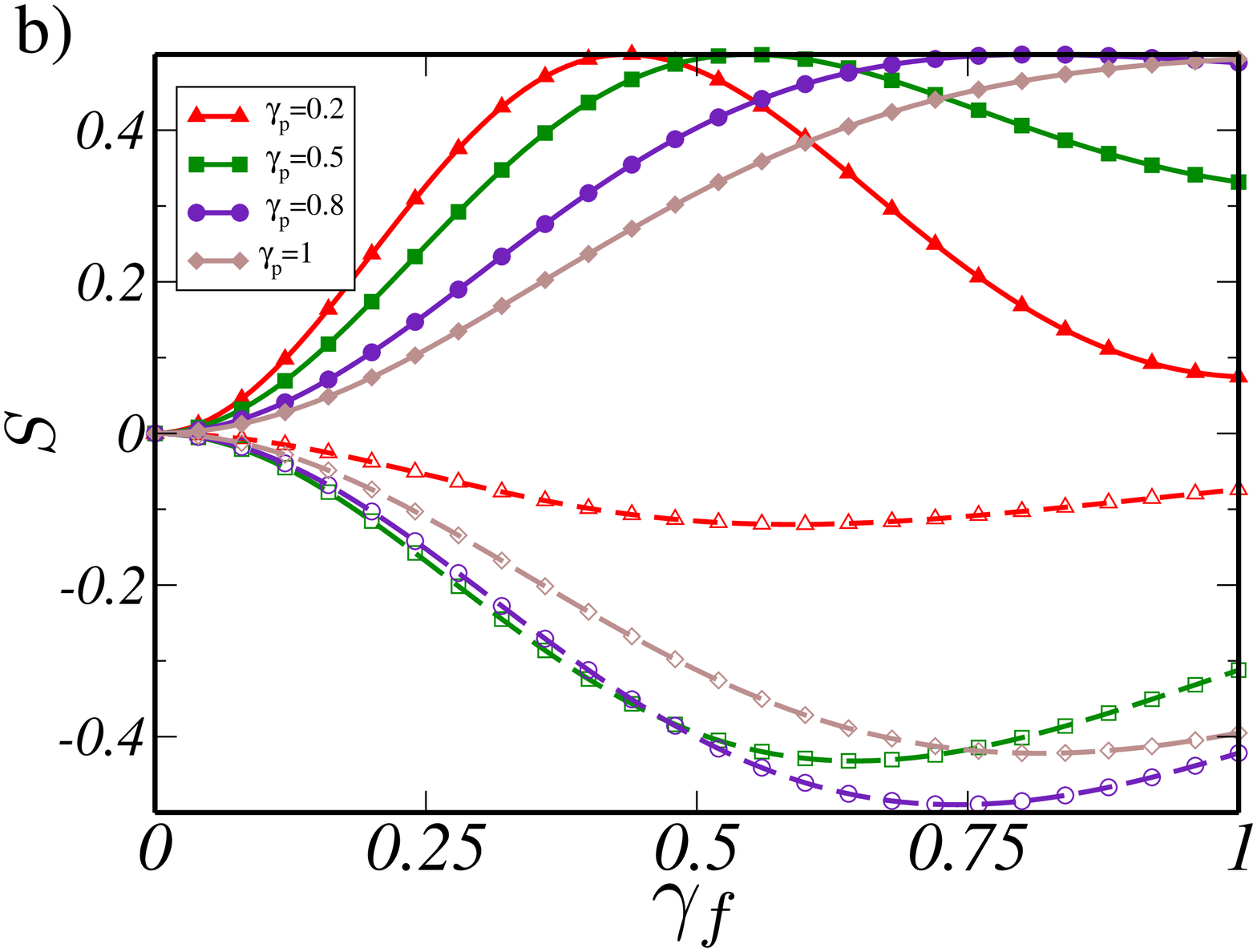}
\caption{(Color online) 
The noise power S in units of $e^{3}V/(2h)$ versus spin-flipping term in a spin-bias configuration $V_1=V_3=0$ and $V_2=V_4=V$. 
a) The cross-correlation functions $S_{3,1}=S_{2,4}$ (dashed blue line) and $S_{2,1}=S_{3,4}$ (solid lines) for currents flowing between the top $l=2,3$ and bottom $l=1,4$ terminals.  
b) The cross-correlation function $S_{2,3}$ (dashed lines with open symbols) and the auto-correlation functions $S_{2,2}=S_{3,3}$ (solid lines with solid symbols) for currents flowing through the top $l=2,3$ terminals. Different colors and symbols correspond to different values of $\gamma_{p}$.
 } 
\label{fig3}
\end{figure}

The corresponding behavior of the noise power is shown in Fig.~\ref{fig3}. 

In the upper panel, Fig.~\ref{fig3} a), we show the cross-correlation function for currents flowing through top and bottom terminals. 
We see that 
the cross-correlation function for currents flowing through
the terminals at the same voltage, see Eq.~(\ref{ntc}), 
is exactly zero, $S_{3,1}=S_{2,4}=0$. 
Instead, the cross-correlation function for currents flowing through the terminals biased differently,
$S_{2,1}=S_{3,4}$, vanishes only for $\gamma_f=0$. 
The vanishing cross-correlator in the latter case is due to the fact that currents flowing in all the terminals do not fluctuate at $\gamma_{f} = 0$, see Fig.~\ref{fig3}.

The relevant components of the correlation function for currents flowing
between the top  terminals are shown in Fig.~\ref{fig3} b), where we show $S_{3,3}= S_{2,2}$ and $S_{2,3}$. The corresponding ones for the
bottom terminals can be inferred from the top ones by noticing that $S_{1,4}=S_{2,3}$, $S_{3,3}=S_{1,1}$ and $S_{4,4}=S_{2,2}$. 
Let us begin analyzing $S_{23}$, which is 
shown in dashed lines in Fig.~\ref{fig3} b). It is zero at $\gamma_{f} =0$, since, as mentioned above, in this case the currents are non-fluctuating.
As $\gamma_f$ is turned on, the current through the QPC increases, implying an increasing noise power $S_{2,3}$.

\subsubsection{The Pauli peak}

\begin{figure}
 \centering
 \includegraphics[width=9.0cm]{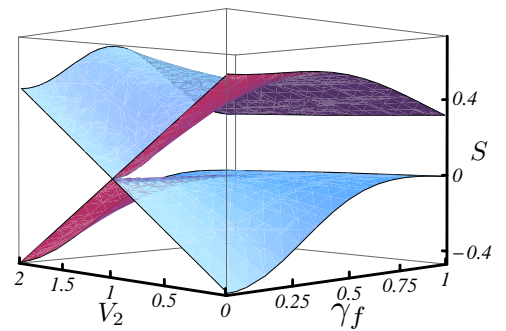}
 \caption{
 (Color online) The cross-correlation function $S_{1,3}$ (the lower sheet) and the auto-correlation function $S_{3,3}$ (the upper sheet) both in units of $e^{3}V/(2h)$ are shown as function of both the spin-flip rate $\gamma_{f}$ and the voltage $V_{2}$. The latter quantity is normalized by $V_{4} = V$, hence, $V_2=V_4$  ($\Delta V=0$) corresponds to $V_2=1$ in the figure.The spin-preserving rate is $\gamma_{p} = 0.5$. The voltages $V_{1} = V_{3} = 0$. 
} 
\label{fig4}
\end{figure}

The electron flows emanated by the metallic contacts $2$ and $4$ are not fluctuating at zero temperature. 
They can fluctuate only after the electron reflections and transmissions  that take place at  the QPC.
In the absence of spin-flip tunneling in the bias configuration defined in Eq.~(\ref{ntc}), electrons with the same spin injected from terminals $2$ and $4$ collide at the QPCs and are backscattered into the terminals $1$ and $3$. 
Thus, if both incoming channels, from the contacts $2$ and $4$ are fully filled up to the same level, that happens at $V_{2} = V_{4}$, then both 
outgoing channels, into contacts $1$ and $3$, will be also fully filled to the same level. 
This is because the Pauli exclusion principle forces the electrons colliding at the QPC to go 
 to different outputs (see, e.g., Refs.~\onlinecite{Blanter:2000wi,Feve:2008im} for a more detailed discussion). 
As a consequence, not only the incoming currents $I_2$ and $I_4$, but also the outgoing ones, $I_{1}$ and $I_{3}$ (if $V_1=V_3=0$) will be non-fluctuating. 
Therefore, for $\gamma_{f}=0$ we have $S_{1,1} = S_{3,3} = 0$ as it is clear in  Fig.~\ref{fig3} b). 
The cross-correlator of non-fluctuating currents is trivially zero, $S_{1,3} =0$. 
If two incoming channels are filled up to different levels, $\Delta V \equiv V_{2} - V_{4} \ne 0$, then the excess flow from the contact with higher level will be scattered between two outgoing channels. 
Since one particle cannot be scattered to both outputs simultaneously, the outgoing streams become fluctuating, $S_{1,1} > 0$, $S_{3,3} > 0$, and $S_{1,3} < 0$, see Fig.~\ref{fig4}. 
These fluctuations are referred to as the shot noise appeared due to indivisibility of carriers.\cite{Blanter:2000wi} 

The spin-flip processes, $\gamma_{f} \ne 0$, open additional scattering channels. In this case, electrons can be  scattered  to   the  terminals $2$ and $4$, which causes current fluctuations even for  $V_{2} = V_{4}$. 
The reason for these current fluctuations is a smaller number of (completely and equally filled) incoming channels in comparison to the number of outgoing channels. 
In fact, spin-flip processes opens the total number of four outgoing channels, while the incoming ones are just two.
It is, however, remarkable that, even in the presence of spin-flip processes the cross-correlator $S_{13} = 0$,  see the plot in dashed lines of Fig.~\ref{fig3} a). 
The vanishing value of the latter cross-correlator for $\gamma_{f}\ne 0$ and $V_{2}=V_{4}$ can be explained as  due to the cancellation of positive two-particle contributions, when electrons from $2$ an $4$ attempt to enter the same terminal $1$ or $3$, and negative single-particle contribution to noise, when each electron is scattered independently. 
For a more detailed discussion, see Ref.~\onlinecite{Moskalets:2011jx}.

In Fig.~\ref{fig4} we show $S_{1,3}$ and $S_{3,3}$ as a function of both the rate of spin-flip tunneling $\gamma_{f}$ and a bias difference $\Delta V = V_{2}-V_{4}$. 
Both correlators are suppressed for $\Delta V=0$. 
Since this suppression is due to the fermionic correlations arising between colliding electrons, we use the names of the Pauli peak for $S_{1,3}$ and the Pauli dip for $S_{3,3}$ as functions of $\Delta V$. 
This peak/dip structure is clearly visible at a small rate of spin-flip scattering, $\gamma_{f} \to 0$, in Fig.~\ref{fig4} . 
While the cross-correlator is suppressed down to zero, the auto-correlator's dip depends on the rate of spin-flip processes $\gamma_{f}$.   
To show this explicitly we use the scattering matrix approach \cite{Buttiker:1992vr,Moskalets:2011cw}, and calculate $S_{3,3}$ analytically at $V_{2} = V_{4} = V$ and $V_{1}=V_{3}=0$~: 

\begin{eqnarray}
& &S_{3,3} = \frac{e^{3}  }{h } \left\{ T_{3,1} \left[|V_{2}|  T_{3,2} +  |V_{4}| T_{3,4} \right]+ |V_{2} - V_{4}| T_{3,2} T_{3,4}   \right\}\,,
\nonumber \\
\label{sg} \\
& & S_{3,1} = - \frac{e^{3}  }{h }  |V_{2} - V_{4}| T_{3,2} T_{3,4}  \,,
\nonumber 
\end{eqnarray}
\ \\
\noindent
where the transmission functions $T_{l,l^{\prime}}$ are presented in Appendix ~\ref{tf}. It becomes apparent from the previous expression that $S_{1,3}=0$ at $V_{2}=V_{4}$. 
For  $\gamma_{f} \to 0$ and $V_{2}=V_{4}=V$ we find the following behavior of $S_{3,3}$ in the leading order in the spin-flip rate,
\begin{eqnarray}
S_{3,3} =  \frac{e^{3} V }{h } \frac{4\gamma_f^2}{(1+\gamma_p^2)^{2}} + {\cal O}\left( \gamma_{f}^{4} \right) \,.
\label{sgsm}
\end{eqnarray}
\ \\
\noindent 
Thus the gap between the maximum of $S_{13}$ and the minimum of $S_{33}$ would unambiguously demonstrate an actual presence of spin-flip scattering at the QPC allowed by the symmetry of helical states. 

The numerical calculations confirm that the  above relations, Eqs.~(\ref{sg}) and (\ref{sgsm}),   completely agree with the results of the Green's function approach, see Eqs.~(\ref{noise}) and (\ref{T312}), (\ref{T34}). 

The Pauli peak in the cross-correlation function (as a function of a bias asymmetry $\Delta V$) is robust and is the direct consequence of the fermionic nature of carriers. 
The gap appearing between the cross- and auto-correlation functions at both $V_{2}=V_{4}$ and $\gamma_{f} \ne 0$  is a direct consequence of the helical nature of the  edge states. 
Another  peculiar manifestation of the helical nature of edge states has been  recently predicted in Ref.~\onlinecite{Edge:2013ed} under the name of $Z_{2}$ peak. 
This feature manifests itself in the dependence of the current cross-correlator on the external magnetic field $B$ as a peak  at $B=0$. The origin is  an exact cancellation of two different components of the noise, the partition noise and the exchange noise. 
Such cancellation is a consequence of the time-reversal symmetry of the helical state. 
It is removed as soon as it is broken by a small magnetic field $B$, in which case the cross-correlation function becomes finite and negative.

For $\gamma_{f}=0$, our setup of helical states can be effectively decoupled into two overlapping but independent sets of chiral edge states connected in pairs by the QPC.  
In this case, the Pauli peak and dip discussed above  is analogous  to the one discussed \cite{Olkhovskaya:2008en} and measured \cite{Bocquillon:2013dp} for electrons emitted with the same rate by two single-electron sources into a  chiral edge state of a two-dimensional electron gas in the quantum Hall effect  coupled by a  QPC. 
The difference  between  emission times $\Delta\tau$ of single-electron sources plays the same role as the voltage difference $\Delta V = V_{2} - V_{4}$ in the present case.  
For  $\Delta\tau=0$ electrons collide at the QPC and due to the Pauli principle they  are necessarily scattered to different outputs. 
Hence, the outgoing currents are noiseless. 
In contrast, if two electrons pass the QPC at different times, $\Delta\tau\ne 0$ they are scattered independently. This results in the shot noise  (a positive auto-correlator and a negative cross-correlator for outgoing currents).   
Instead, for $\gamma_{f} \ne 0$, the behavior of the noise power in our setup with helical states is analogous to an  electronic circuit with chiral states as waveguides having the number of outgoing states exceeding the number of populated incoming states. 
In particular, in Ref.~\onlinecite{Moskalets:2011jx} it was shown that  for the synchronized emission of electrons in two incoming channels (the analogue of $\Delta V=0$ in our setup) the cross-correlator vanishes  while the auto-correlator remains finite
as in our case.

\subsection{Two point contacts}
In the case of an interferometer with two QPC, the current exhibits oscillations when plotted against the gate potentials. The relevant transmission functions can be calculated from the retarded
Green's function and the hybridization function presented in Appendix \ref{apc}. For completeness, we present the explicit expression for these functions in Appendix \ref{td2}.
The different interference processes are characterized by two phases defined in Eq.(\ref{phases}), the Fabry-P\'{e}rot phase $\phi_{FP}$, which is symmetric in gate voltages, and the Mach-Zehnder phase $\phi_{MZ}$, which is antisymmetric in gate voltages. The first phase is associated with the spin-preserving tunneling when a particle traverses  the arms of the interferometer in the same direction (clockwise or anticlockwise ) like in the Fabry-P\'{e}rot interferometer.  Instead, the second phase is associated with the spin-flip tunneling, which forces an electron to traverse the interferometer only once similarly to what happens in the Mach-Zehnder  interferometer. 

At this point, we would like to further justify the choice of the name $MZ$  for the phase that is antisymmetric in the gate voltages. Notice that it differs from the one used in Refs.~\onlinecite{Dolcini:2011dp,Citro:2011fw,romeo,Ferraro:2013du}, where it  is referred to as the ''Aharonov-Bohm'' ($AB$) phase. 
Since the latter suggests the phase induced by the vector potential associated to a
magnetic flux penetrating the system, \cite{Webb:1985tu} we prefer to avoid that denomination in the present system, which preserves time-reversal invariance. 
There is literature on the Aharonov-Bohm effect induced by the scalar potential, \cite{Washburn:1987kn,VanOudenaarden:1998ve, deVegvar:1989fe} which could eventually justify the use of that term in the present context.
 However, it is not clear in those systems whether the effect is due to the extra phase acquired by electrons according to the Aharonov-Bohm mechanism\cite{Aharonov:1959wh,Aharonov:1961ub} or simply due to the change of the electron trajectories\cite{Washburn:1987kn} or the electron density.\cite{deVegvar:1989fe} For this reason, we
 find it more appropriate to use
 the name Mach-Zehnder  with the aim of emphasizing that the electron traverse the interferometer's arms in the same direction. \cite{Ji:2003ck} Generally both phases $\phi_{MZ}$ and $\phi_{FP}$ are involved into the same interference process.

\subsubsection{Current}

\begin{figure}
 \centering
 \includegraphics[width=8cm]{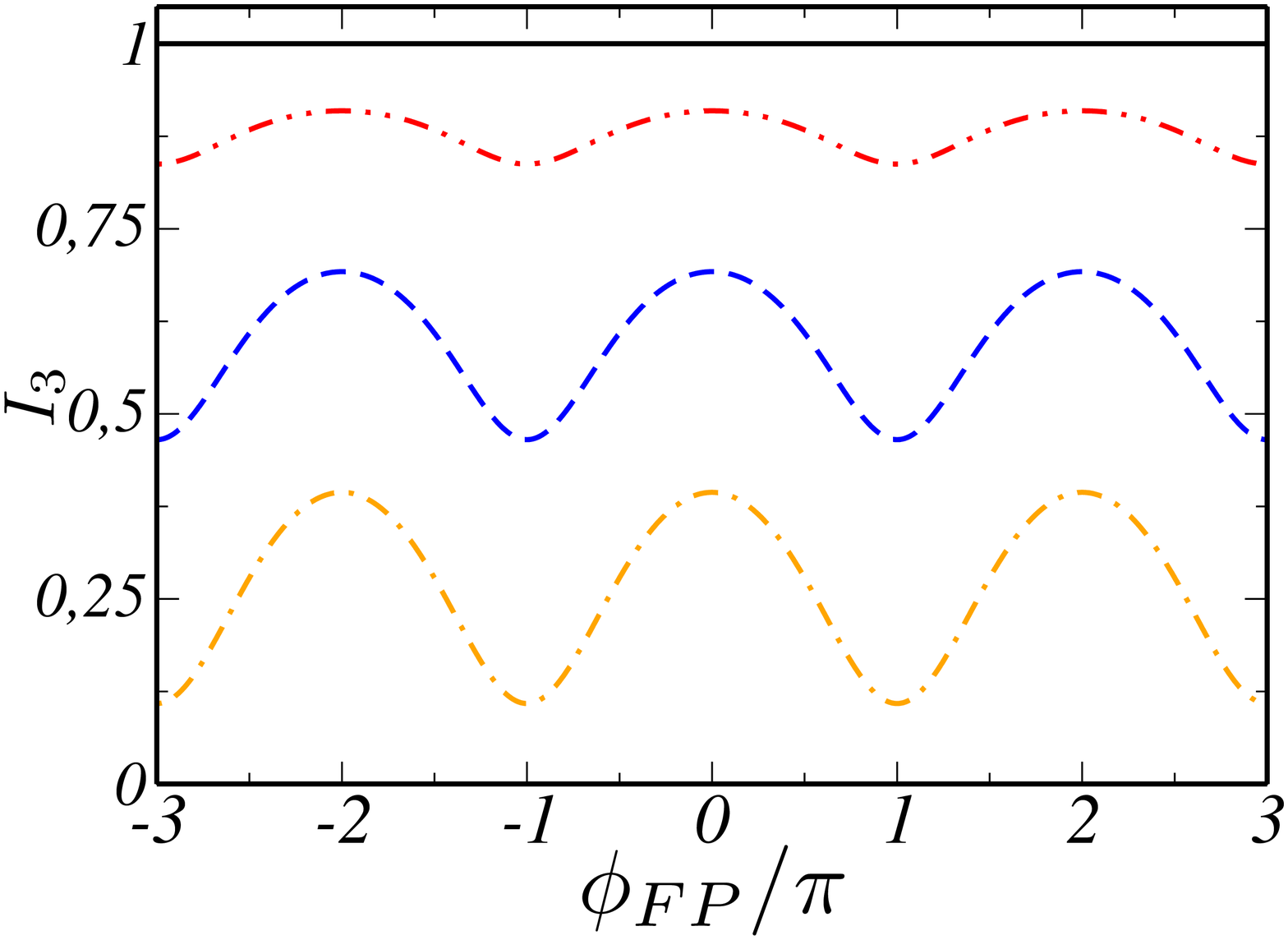}\vspace{0.9cm}\\
 \includegraphics[width=8cm]{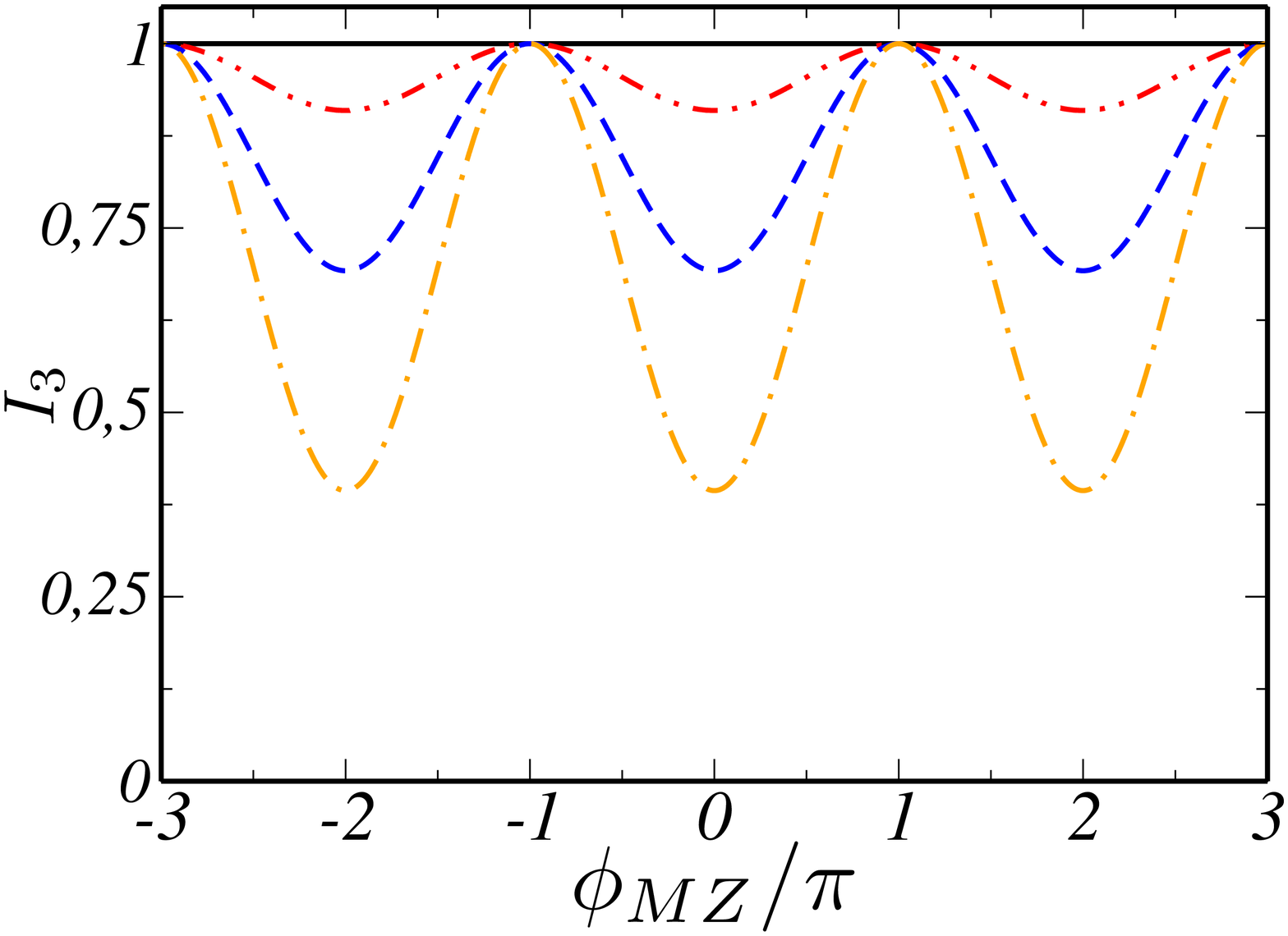}
 \caption{(Color online) Current oscillations in terminal 3 versus the Fabry-P\'{e}rot phase $\phi_{FP}=\frac{eL(V_{gT}+V_{gB})}{\hbar v_F}$ and versus the Mach-Zehnder phase 
 $\phi_{MZ}=\frac{eL(V_{gB}-V_{gT})}{\hbar v_F}$. The current is given in units of  $e^{2}V/h$. We set $\gamma_{p}=0.2$ and $\gamma_{f}=0$ (black solid curve), 
 $\gamma_{f}=0.1$ (red dashed- double dotted curve),  $\gamma_{f}=0.2$ (blue dotted curve) and $\gamma_{f}=0.5$ (orange dashed-dotted curve).}
\label{fig5}
\end{figure}

In Fig.~\ref{fig5}, we show the behavior of the current at the Fermi energy in terminal $l=3$ versus the  two phases for the configuration $V_1=V_3=0$ and $V_2=V_4=V$, and different values of the spin-flipping tunneling. As in the case of a single QPC, the different currents are related as $I_1=I_3$, $I_2=I_4$ and $I_3=-I_4$.

Interestingly, for this configuration of voltages, the interference is not effective for
$\gamma_f=0$. Thus, for vanishing spin-flip tunneling, the current in terminals $l=1,3$ is just 
$I_3^{0}=I_1^{0}$.  
As in the case of a single QPC, the reason for this behavior is the Pauli exclusion principle according to which both outgoing channels have to be equally populated if two incoming channels are equally populated. Therefore, the outgoing currents are not sensitive to gate voltages at this bias configuration.

 For $\gamma_f \neq 0$, 
the current through the interferometer becomes sensitive to gate voltages. 
Moreover, 
the  two interference processes compete.  
In the upper panel we show that for vanishing $\phi_{MZ}$, interference effects
systematically decrease the conductance through the terminal $l=3$ bellow the conductance quantum $e^2/h$. As functions of $\phi_{FP}$ it presents maxima (minima) at 
$\phi_{FP}= 2m \pi$, ($\phi_{FP}= (2 m+1) \pi $), with $m$ integer. Instead, for $\phi_{FP}=0$, the $l=3$ conductance  as a function of $\phi_{MZ}$
achieves maxima equal to the conductance quantum $e^2/h$ 
for $\phi_{MZ}= (2 m+1) \pi$ and minima for $\phi_{MZ}=2m \pi$ which become deeper as $\gamma_f$ increases.

\subsubsection{Noise}

\begin{figure}
 \centering
 \includegraphics[width=8cm]{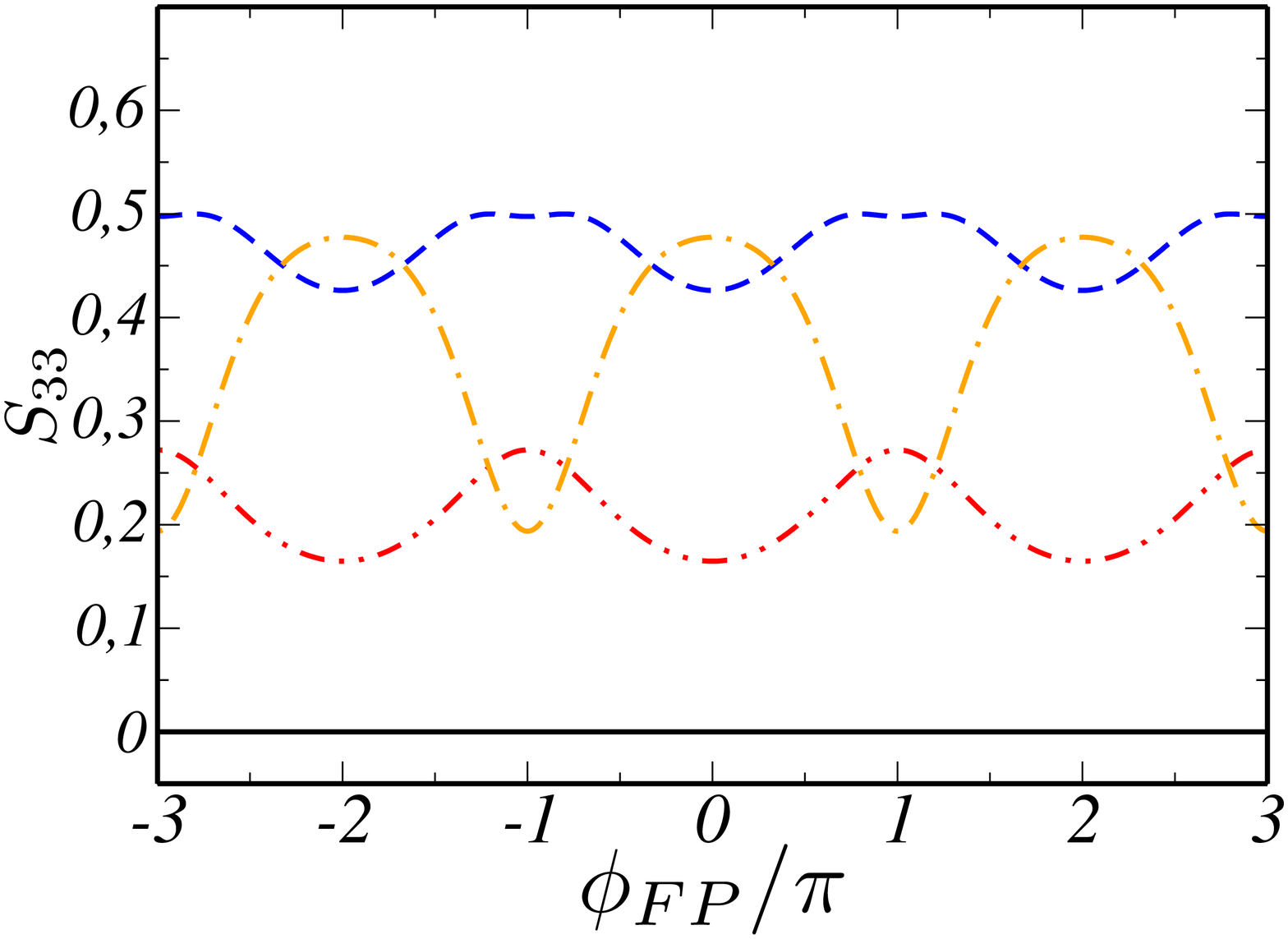}\vspace{0.9cm}\\
 \includegraphics[width=8cm]{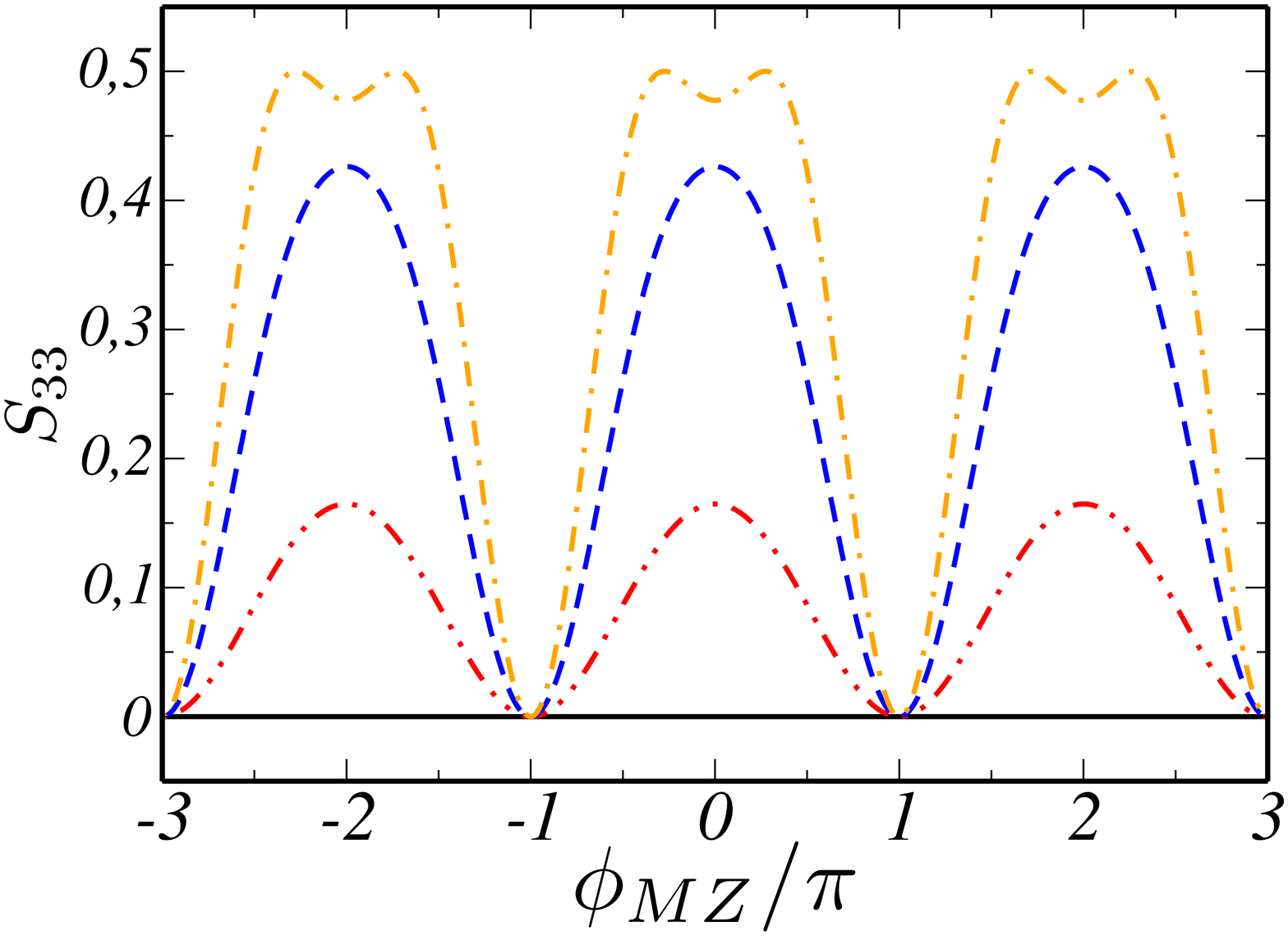}
 \caption{(Color online) Shot-Noise oscillations of terminal $l=3$ versus the Fabry-P\'{e}rot phase $\phi_{FP}=\frac{eL(V_{gT}+V_{gB})}{\hbar v_F}$ ($\phi_{MZ}=0$) and versus the Mach-Zehnder phase $\phi_{MZ}=\frac{eL(V_{gB}-V_{gT})}{\hbar v_F}$ ($\phi_{FP}=0$), for a spin-bias configuration $V_1=V_3=0$ and $V_2=V_4=V$. The noise power is given in units of $e^{3}V/(2h)$.  We set $\gamma_{p}=0.2$ and $\gamma_{f}=0$ (black solid curve), $\gamma_{f}=0.1$ (red dashed- double dotted curve),  $\gamma_{f}=0.2$ (blue dotted curve) and $\gamma_{f}=0.5$ (orange dashed-dotted curve).}
\label{fig6}
\end{figure}

The corresponding behavior of the noise power at the terminal $l=3$ is shown in Fig.~\ref{fig6}. 

The perfect transmission through this terminal observed in Fig.~\ref{fig5} for 
$\gamma_f=0$ is related to a vanishing noise, $S_{3,3}=0$. As this tunneling parameter is switched on the noise power displays oscillations as a function of
$\phi_{MZ}$ and $\phi_{FP}$. In Fig.~\ref{fig6}a we show the dependence on $\phi_{FP}$  of the noise power $S_{3,3}$ for $\phi_{MZ}=0$. 
We identify a value $\gamma_f^*$ such that for
small $\gamma_f <  \gamma_f^*$, the noise power 
increases and has maxima (minima) at the phases where the conductance has minima (maxima), i.e. for $\phi_{FP}= (2 m+1) \pi $,  ($\phi_{FP}= 2m \pi$) with $m$ integer.
For the parameters of the Fig. $\gamma_f^* \sim 0.25$. This value of 
$\gamma_f $  corresponds to the one for which the conductance of this terminal  achieves the value $G_3=I_3/V=e^2/(2h)$ for some values of $\phi_{FP}$. 
The corresponding noise power at these points is the maximum possible value $S_{3,3} =e^3 V/(2h)$ and the noise power turn to have local minima for $\phi_{FP}=(2 m+1) \pi$. For large enough $\gamma_f$, 
$G_3<e^2/(2h),\; \forall \phi_{FP}$ and the noise power follows the pattern of maxima (for $\phi_{FP}= 2 m \pi $) and minima (for $\phi_{FP}= (2 m+1) \pi$) of the conductance.
Fig.~\ref{fig6}b shows the dependence on $\phi_{MZ}$  of the noise power $S_{3,3}$ for $\phi_{FP}=0$. In this case, the noise vanishes for $\phi_{MZ}= (2 m+1) \pi$, for which 
$G_3= e^2/h$. For $\gamma_f < \gamma_f^* $ the noise power displays maxima at $\phi_{MZ}= 2 m \pi$, but this maxima turn to local minima for $\gamma_f \geq \gamma_f^*$ as a consequence
of the fact that $G_3< e^2/(2h)$ for these values.

\begin{figure}
 \centering
 \includegraphics[width=8cm]{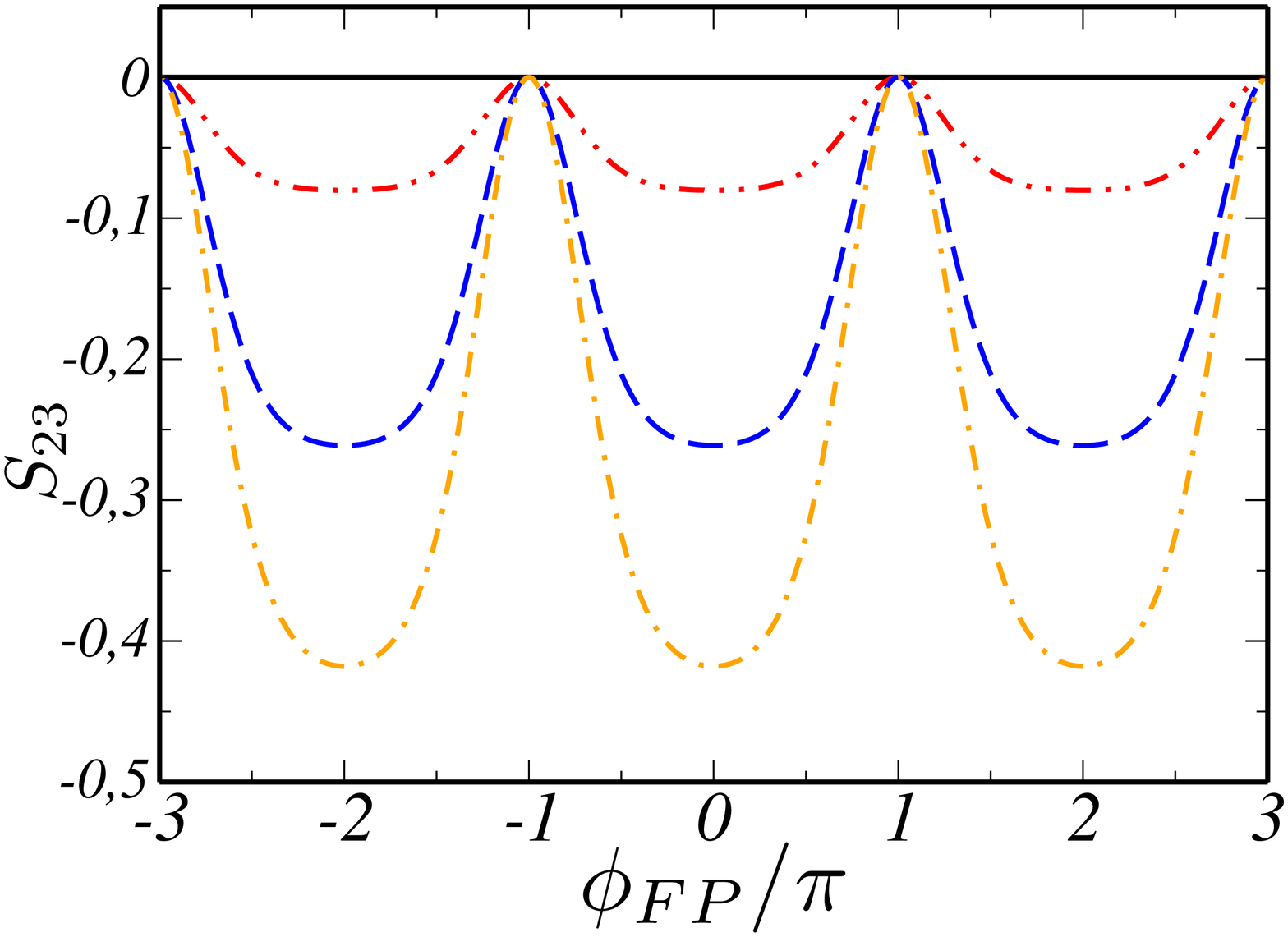}\vspace{0.9cm} \\
 \includegraphics[width=8cm]{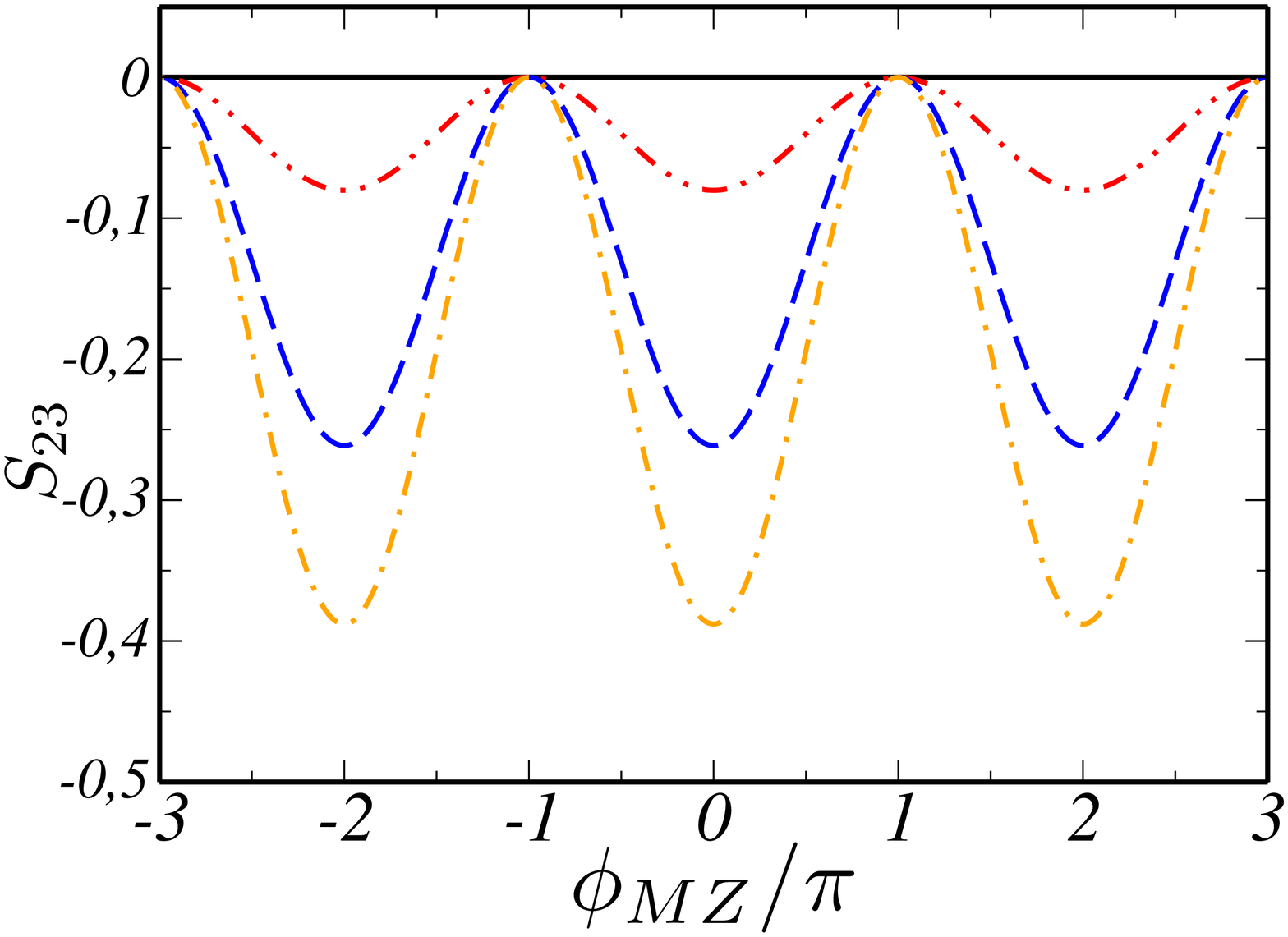}
 \caption{(Color online) Shot-Noise oscillations between terminals $l=2$ and $l=3$ versus the Fabry-P\'{e}rot phase $\phi_{FP}=\frac{eL(V_{gT}+V_{gB})}{\hbar v_F}$ and versus the Mach-Zehnder phase $\phi_{MZ}=\frac{eL(V_{gB}-V_{gT})}{\hbar v_F}$, for a spin-bias configuration $V_1=V_3=0$ and $V_2=V_4=V$. The noise power is given in units of $e^{3}V/(2h)$. We set $\gamma_{p}=0.2$ and $\gamma_{f}=0$ (black solid curve), $\gamma_{f}=0.1$ (red dashed- double dotted curve),  $\gamma_{f}=0.2$ (blue dotted curve) and $\gamma_{f}=0.5$ (orange dashed-dotted curve). Note that the cross-correlator of bottom terminals $S_{1,4}=S_{2,3}$. }
\label{fig7}
\end{figure}

The cross correlation between the currents through the  top terminals are shown in Fig.~\ref{fig7}. 

\begin{figure}
 \centering
 \includegraphics[width=8cm]{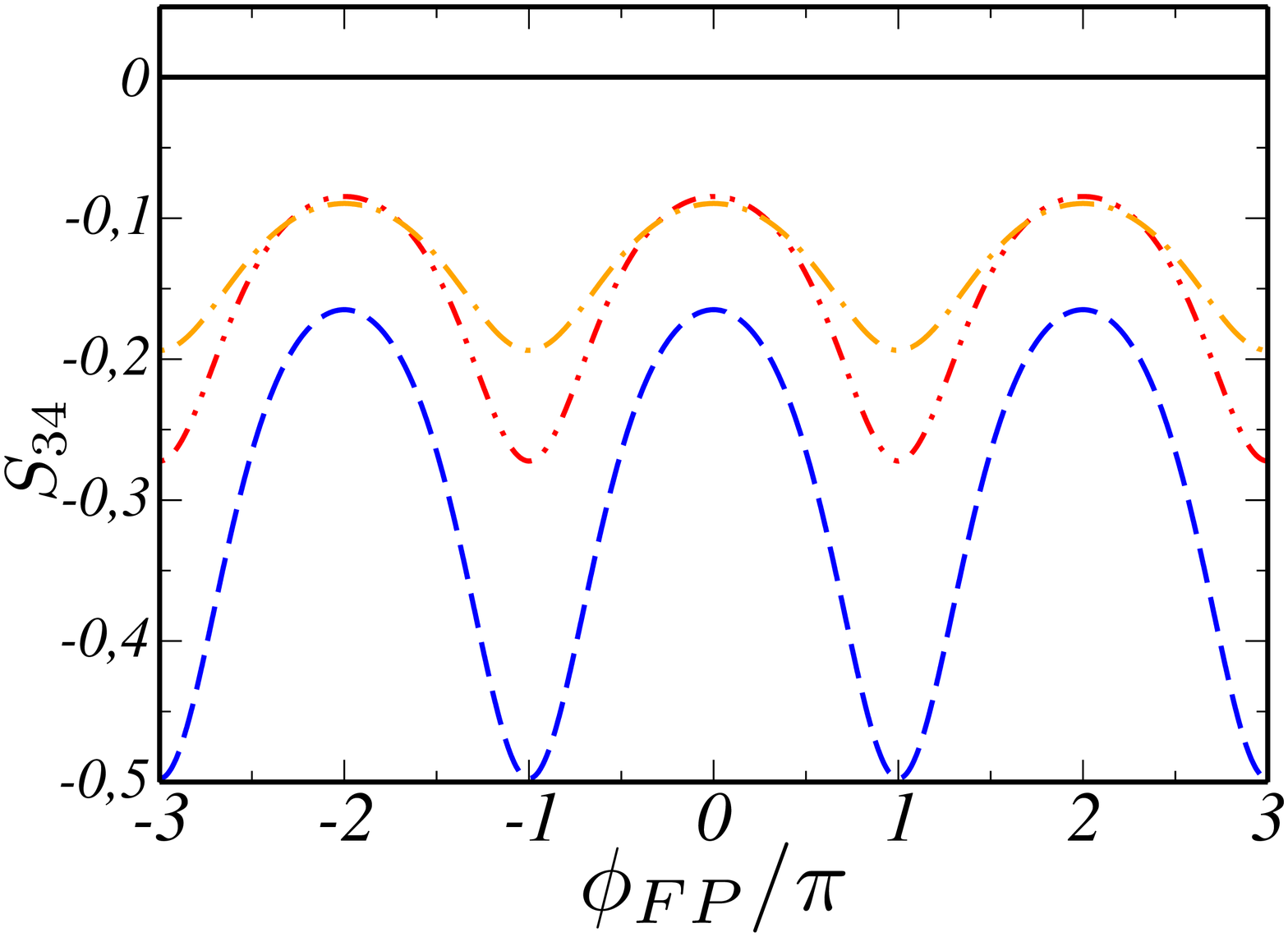}\vspace{0.9cm}\\
  \includegraphics[width=8.cm]{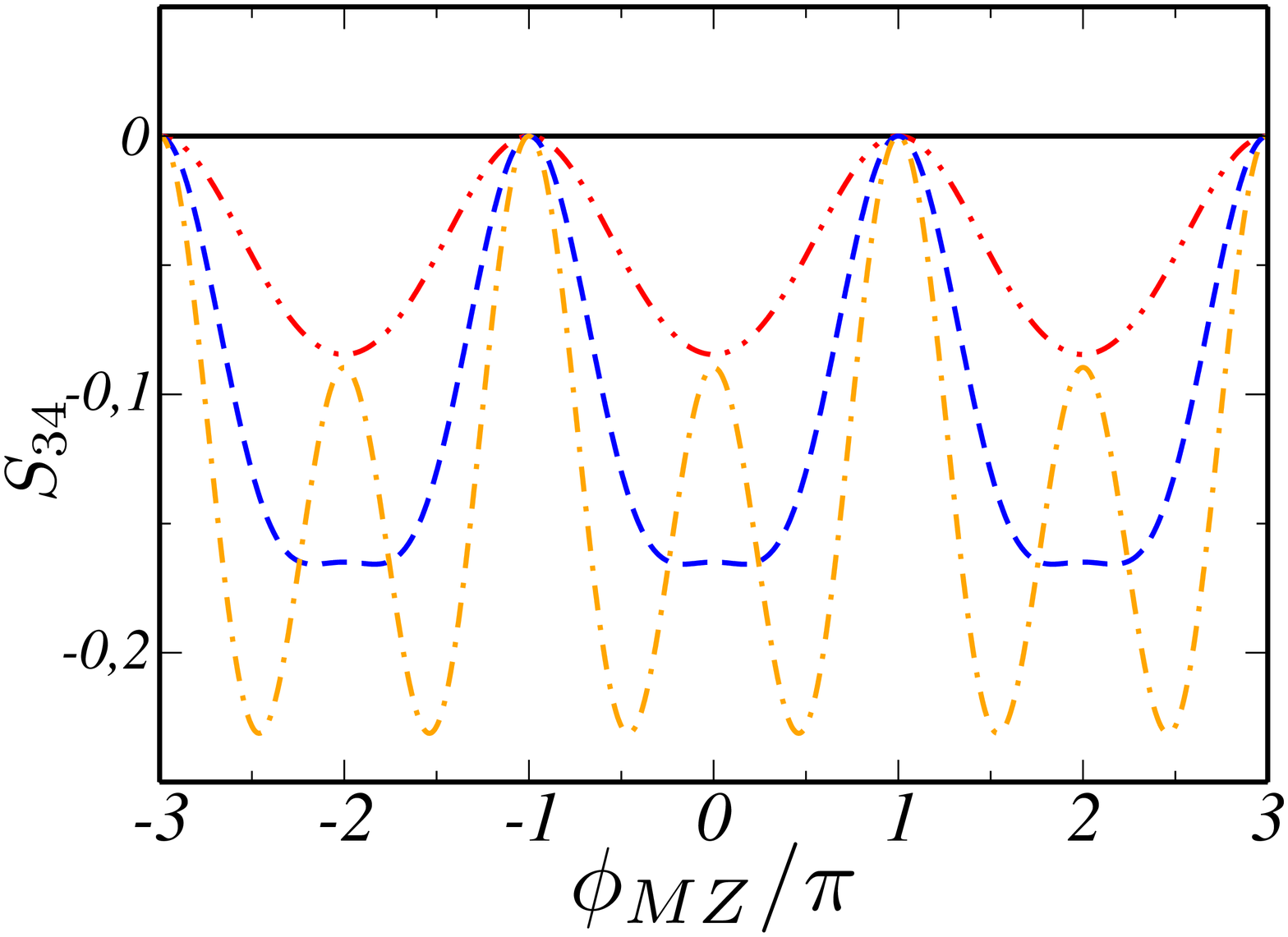}
 \caption{(Color online) Shot-Noise oscillations between top $l=3$ and bottom $l=4$ terminals versus the Fabry-P\'{e}rot phase $\phi_{FP}=\frac{eL(V_{gT}+V_{gB})}{\hbar v_F}$, 
 and versus the Mach-Zehnder phase $\phi_{MZ}=\frac{eL(V_{gB}-V_{gT})}{\hbar v_F}$,
 for a spin-bias configuration $V_1=V_3=0$ and $V_2=V_4=V$. We show the oscillations for distinct values of spin-flipping term $\gamma_{f}=0$ (black solid curve),$\gamma_{f}=0.1$ (red dashed- double dotted curve),$\gamma_{f}=0.2$ (blue dotted curve),$\gamma_{f}=0.5$ (orange dashed-dotted curve). The noise power is given in units of $e^{3}V/(2h)$. We set $\phi_{MZ}=0$ and $\gamma_{p}=0.2$. For the rest of top-bottom correlators we have that $S_{1,3}=S_{2,4}=0$ for any value of $\gamma_{f}$ and $S_{1,2}=S_{3,4}$.}
\label{fig8}
\end{figure}

Here we observe that the minima (maxima) of $|S_{2,3}|$ always occur at $\phi_{FP(MZ)}=2m\pi$. 
The maxima do not reach the bound $e^3 V/(2 h)$. 
Finally, in Fig.~\ref{fig8} the top-bottom correlation function $S_{3,4}$ is showed.
This noise is zero just for $\gamma_{f}=0$, in contrast of $S_{1,3}$ (or $S_{2,4}$) which vanish for any value of $\gamma_f$. 
The absolute  value of the top-bottom noise correlations for  $\phi_{MZ}=0$, shown in the upper panel of  Fig.~\ref{fig8}, has minima (maxima) for $\phi_{FP}=2m \pi$ ($ \phi_{FP}= (2m+1) \pi$). 
For $\phi_{FP}=0$ this quantity presents a more complicated structure as a function  of $\phi_{MZ}$ as the tunneling parameter $\gamma_f$ varies. 
For any value of  this parameter, it vanishes at   $\phi_{MZ}= (2m+1)\pi$. 
 For $\gamma_f < \gamma_f^*$ the absolute value $|S_{2,4}|$ has maxima at $\phi_{MZ}= 2m \pi$, while for $\gamma_f \geq \gamma_f^*$ it has local minima for these values of $\phi_{MZ}$.
Unlike other cases, $|S_{2,4}|$ does not reach the upper bound $e^3 V/(2 h)$ for any phase $\phi_{MZ}$.

\section{Summary and conclusions}
\label{concl}

We have presented a formal treatment based on non-equilibrium Green's function formalism to study the transport properties of interferometers of helical edge sates in bar geometries.
We have derived expressions for the currents and we have defined  transmission functions. The latter are the building blocks of the scattering matrix approach. In setups consisting of a small system
connected to two or more particle reservoirs, the relation between these two formalisms is well known since the work by Fisher and Lee \cite{fisher-lee} for stationary transport. In Ref. \onlinecite{rel} the
suitable generalization to systems with ac driving was carefully analyzed. The case of currents flowing through edge states is special in the sense  that the latter constitute reservoirs that support chiral
currents. The contacts to metallic terminals just set the proper imbalance between them giving rise to a net current. In the present work, we have presented the generalization of Fisher and Lee's  relation between the two formalisms for edge states currents (see sections \ref{gtf}, \ref{rsm}). We have also presented the expressions for the noise power in the Green's function formalism.

We have used these formal tools to analyze the transport properties of  helical electronic circuits containing one or two quantum point contacts, inducing tunneling and scattering processes between the different edge states.
Typically this type of effects are expected to take place preserving or flipping spin. 
We have focused on identifying the features induced in the transport properties (currents and current-current correlation functions) originated in the spin-flip scattering processes at the QPC allowed by the symmetry of  helical states.
Importantly, for the case of a single QPC, we have identified a peculiar feature, the Pauli peak-dip structure in the current-current correlation functions, see Fig.~\ref{fig4}, which allows to identify the presence or absence of a spin-flip scattering. 
For two QPC, following Refs.~\onlinecite{Dolcini:2011dp,Citro:2011fw,romeo,Ferraro:2013du} we have assumed in the present work that applying gates within a finite region of the top and bottom
edges induce additional potentials $V_{gT}$ and $V_{gB}$ within the helical state. 
These additional potentials in turn change the electron phase, which yields a rich structure of currents and current-current correlation functions, with patterns of maxima and minima. This structure is the consequence of competing interference processes and has two important experimental outcomes. On one hand,  the necessary condition for these features to exist it a non-vanishing value of the spin-flip tunneling. Thus,  the observation of these patterns is an important signature to identify and quantify the
relevance of the spin-flip tunneling. On the other hand, their observation would enable to verify the idea that the electron phase in helical edge states can be locally controlled by
this type of gates. That would be  an important breakthrough, which would open a wide research avenue in the usage of solid state linear electronics for efficient quantum computation.
 In fact, with such phase shifters, quantum point contacts as wave splitters, single electron sources,\cite{Feve:2007jx,Blumenthal:2007ho} and metallic contacts as detectors, all
the  necessary elements for quantum computation would be available  in full analogy with quantum optics.\cite{Knill:2001is}

\section{Acknowledgements}
LA thanks A-P Jauho for useful conversations. LA and BR thank support from CONICET, MINCyT and UBACYT, Argentina.

\appendix
\section{Evaluating the Dyson's equation for the Green's functions}\label{apa}
The equation of motion (\ref{eqofmotion}) leads to the Dyson's equation (DE) for the retarded Green's functions (\ref{ret}). The four elements with edge indices 
$\alpha \sigma$ and
$\overline{\alpha} \overline{\sigma}$ can be organized in $2 \times 2$ matrices as follows
\ba \label{apadyret1}
G^{r}(x,x',\omega) &= &g^{0,r}(x,x',\omega)+\sum_{j,j^{\prime}} G^{r}(x,x_{j},\omega) \times \nonumber \\
& & \Sigma^{0,r}(x_j,x_{j^{\prime}},\omega) g^{0,r}(x_{j^{\prime}},x',\omega),
\ea
with
\be
G^{r}(x,x',\omega)=\left( \begin{array}{cc}
G_{\alpha \sigma,\alpha \sigma}^{r}(x,x',\omega) & G_{\alpha \sigma,\overline{\alpha} \overline{\sigma}}^{r}(x,x',\omega)  \\
G_{\overline{\alpha} \overline{\sigma},\alpha \sigma}^{r}(x,x',\omega) & G_{\overline{\alpha} \overline{\sigma},\overline{\alpha} \overline{\sigma}}^{r}(x,x',\omega)  \\  \end{array} \right),
\ee
and 
\be
g^{0,r}(x,x',\omega)=\left( \begin{array}{cc}
g_{\alpha \sigma}^{0,r}(x,x',\omega) & 0  \\
0 & g_{\overline{\alpha} \overline{\sigma}}^{0,r}(x,x',\omega)  \\  \end{array} \right).
\ee
The elements of the above matrix are the retarded Green's function for the uncoupled edge 
\ba \label{g0r}
g^{0, r}_{\alpha\sigma}(x, x^{\prime},\omega) &=& 
\frac{1}{ \hbar} \int_{-k_0}^{+k_0} dk e^{i k(x- x^{\prime})} g^{0,r}_k(\omega), \nonumber \\
 g^{0,r}_k(\omega) & = & \frac{1}{\omega - (v_{\alpha}\hbar k+eV_{g,\alpha\sigma}) + i\eta}, 
\ea
which in the limit of $k_0 \rightarrow \infty$ results
\ba \label{g0rr}
g^{0, r}_{\alpha\sigma}(x, x^{\prime},\omega) & = &  -\frac{ i}{ v_F \hbar} \Theta\left( s_{\alpha} (x-x^{\prime}) \right) \times \nonumber \\
& &
\exp{\left[i \frac{1}{v_F\hbar}(\omega-eV_{g,\alpha\sigma})  (x- x^{\prime}) \right]},
\ea
where $V_{g,\alpha\sigma}=V_{g,T}$ if $\alpha\sigma=R\uparrow,L\downarrow$ and $V_{g,\alpha\sigma}=V_{g,B}$ if $\alpha\sigma=R\downarrow,L\uparrow$.

The self energy in (\ref{apadyret1}) is defined by back substituting the DE for the Green's functions with indices $\alpha \sigma$,
$\overline{\alpha} \sigma$ and 
$\alpha \sigma$,$\alpha \overline{\sigma}$:
\begin{widetext}
\ba\label{apadyret2}
G_{\alpha \sigma,\alpha \overline{\sigma}}^{r}(x,x^{\prime},\omega)& = & \sum_{j}2\hbar v_F [\gamma_{p}G_{\alpha \sigma,\overline{\alpha} \overline{\sigma}}^{r}(x,x_j,\omega)
+ s_{\alpha} \gamma_{f} G_{\alpha \sigma,\alpha \sigma}^{r}(x,x_j,\omega) ]g_{\alpha\overline{\sigma}}^{0,r}(x_j,x^{\prime},\omega),\nonumber
\ea
and
\ba\label{apadyret3}
G_{\alpha \sigma,\overline{\alpha} \sigma}^{r}(x,x^{\prime},\omega) &= & \sum_{j}2\hbar v_F[\gamma_{p}G_{\alpha \sigma,\alpha \sigma}^{r}(x,x_j,\omega) 
+s_{\overline{\alpha}} \gamma_{f} G_{\alpha \sigma,\overline{\alpha} \overline{\sigma}}^{r}(x,x_j,\omega)]g_{\overline{\alpha}\sigma}^{0,r}(x_j,x^{\prime},\omega).
\nonumber
\ea   
\end{widetext}
The result is 
\begin{widetext}
\ba \label{sigma}
&&\Sigma^{0,r}(x_j,x_{j^{\prime}},\omega)= 4\hbar^2 v_F^2 \times\nonumber\\
&& \left( \begin{array}{cc}
\gamma_{p}^{2} g_{\overline{\alpha} \sigma}^{0,r}(x_j,x_{j^{\prime}},\omega)+\gamma_{f}^{2} g_{\alpha \overline{\sigma}}^{0,r}(x_j,x_{j^{\prime}},\omega) 
& \gamma_{f}\gamma_{p} \left[ s_{\overline{\alpha}} g_{\overline{\alpha} \sigma}^{0,r}(x_j,x_{j^{\prime}},\omega)+s_{\alpha} g_{\alpha \overline{\sigma}}^{0,r}(x_j,x_{j^{\prime}},\omega)\right]  \\
\gamma_{f}\gamma_{p} \left[s_{\overline{\alpha}} g_{\overline{\alpha} \sigma}^{0,r}(x_j,x_{j^{\prime}},\omega)+s_{\alpha} g_{\alpha \overline{\sigma}}^{0,r}(x_j,x_{j^{\prime}},\omega)\right]
& \gamma_{f}^{2} g_{\overline{\alpha} \sigma}^{0,r}(x_j,x_{j^{\prime}},\omega)+\gamma_{p}^{2} g_{\alpha \overline{\sigma}}^{0,r}(x_j,x_{j^{\prime}},\omega) \end{array} \right).
\ea
\end{widetext}

The DEs for the lesser Green's functions can be derived from Eqs. (\ref{apadyret1}), (\ref{apadyret2}) and (\ref{apadyret3}) by recourse to Langreth rules,\cite{jauho} according to which
given a product of retarded Green's functions $A^r B^r$, then $(AB)^<= A^< B^a + A^r B^<$.  
In the case of (\ref{apadyret1}) the result is 
\begin{widetext}
\ba
 G^<(x, x^{\prime},\omega) &=& \sum_{j^{\prime},j^{\prime \prime}=1}^M
\left[ \Lambda^{0,r}(x, x_{j^{\prime}}, \omega) +\delta(x-x_{j^{\prime}})  \right]
g^{0,<}(x_{j^{\prime}}, x_{j^{\prime \prime}},\omega)  [ \delta(x_{j^{\prime \prime}}-x^{\prime}) + 
 \Lambda^{0,a}(x_{j^{\prime \prime}}, x^{\prime}, \omega)  ] \nonumber \\
& & +\sum_{j,j^{\prime}=1}^M
G^r(x, x_j, \omega) \Sigma^{0,<}(x_j, x_{j^{\prime}})  
G^a(x_{j^{\prime}}, x^{\prime}, \omega), \label{dysonl}
\ea
\end{widetext}
with $\Lambda^{0,r}(x, x_{j^{\prime}}, \omega) =
\sum_{j} G^r(x, x_j, \omega) \Sigma^{0,r}(x_j, x_{j^{\prime}},\omega) =
[\Lambda^{0,a}(x_{j^{\prime}}, x, \omega)]^{\dagger}$, while
\ba \label{g0les}
  g_{\alpha\sigma}^{0,<}(x, x^{\prime},\omega) & = &
i f_{\alpha,\sigma}(\omega) \rho^0_{\alpha\sigma}(x, x^{\prime},\omega), \nonumber \\
 \rho^0_{\alpha\sigma}(x, x^{\prime},\omega) & = & i \left[g_{\alpha\sigma}^{0,r}(x, x^{\prime},\omega)-g_{\alpha\sigma}^{0,a}(x,x^{\prime},\omega)\right]\nonumber \\
&  = &
\frac{1}{\hbar} \int dk e^{-i k(x- x^{\prime})} \delta(\omega - \epsilon_{\alpha\sigma}).
\ea
Here $\epsilon_{\alpha\sigma}=s_{\alpha}v_F\hbar k+eV_{g,\alpha\sigma}$, with $s_{\alpha}= \pm 1, \; \alpha=R,L$ and 
$\sigma= \uparrow,  \downarrow$, while 
$ f_{\alpha,\sigma}(\omega)= f(\omega - \mu_{\alpha, \sigma})$, and  $\mu_{\alpha,\sigma}$ is the 
chemical potential of the reservoir {\em from where the electrons are injected}.

\section{Retarded Green's function for a single QPC}

The Dyson equation (\ref{apadyret1}) for a single QPC at $x_1$ can be solved resulting,
\be \label{gr1}
G_{R,\uparrow}^{r}(x, x_1,\omega)=\frac{G_{R,\uparrow}^{0}(x, x_1,\omega)}{1-\Sigma_{R,\uparrow}^{R}(x_1, x_1,\omega)G_{R,\uparrow}^{0}(x_1, x_1,\omega)}.
\ee
Using Eq.(\ref{g0r}) and Eq.(\ref{sigma}) we find
\be
G_{R,\uparrow}^{r}(x ,x_1,\omega)=-\frac{i\Theta(x-x_1)}{v_F \hbar}\frac{e^{i\frac{\omega}{\hbar v_F}(x-x_1)}}{1+\gamma_p^{2}+\gamma_f^2}.
\ee
The other components are calculated from the following relations $G^r_{L,\uparrow}(x,x^{\prime},\omega)= G^r_{R,\uparrow}(x^{\prime}, x,\omega)$
and $G^r_{\alpha, \downarrow}(x,x^{\prime}, \omega)= G^r_{\alpha, \uparrow}(x,x^{\prime}, \omega),\;\; \alpha= L, R$. The hybridization term is $\Gamma_{L\uparrow(R\downarrow)}=4v_F\hbar\gamma_{p(f)}^2$.  
 
\section{Retarded Green's function for two QPCs}
\label{apc}

We consider two point contacts at $x_1$ and $x_2$, with $x_1 < x_2$.In the general case ($\gamma_{p(f)}\neq 0$), the equations for the
two spin species are coupled and the solution for $x>x_2$ of the diagonal for the $R, \uparrow$ movers reads

\begin{widetext}
\ba
&& {\cal G}_{R\uparrow}^r(x,x_1;\omega)=\frac{ie^{\frac{i}{v_F\hbar}(x-x_1)(\omega-eV_{g,T})}}{\hbar v_F\overline{\Delta}} 
\left[-1+\gamma_{p}^{2}+\gamma_{f}^{2}\left(1+2e^{\frac{iL}{v_F\hbar}(eV_{g,T}-eV_{g,B})}\right)\right], \nonumber\\
&& {\cal G}_{R\uparrow}^r(x,x_2;\omega)=-\frac{ie^{\frac{i}{v_F\hbar}(x-x_2)(\omega-eV_{g,T})}}{\hbar v_F\overline{\Delta}}\left[1+\gamma_{f}^{2}+\gamma_{p}^{2}\left(1+2e^{\frac{iL}{v_F\hbar}(2\omega-eV_{g,T}-eV_{g,B})}\right)\right],
\ea
\end{widetext}
with $\overline{\Delta}=4\gamma_{p}^{2}e^{i\frac{L}{\hbar v_F}(2\omega-eV_{gT}-
eV_{gB})}+(1+\gamma_{p}^{2}+\gamma_{f}^{2})^{2}$.
For $x<x_1$, we have
\begin{widetext}
\ba
&& {\cal G}_{L\downarrow}^r(x,x_1;\omega)= -\frac{ie^{-\frac{i}{v_F\hbar}(x-x_1)(\omega-eV_{g,T})}}{\hbar v_F\overline{\Delta}}\left[1+\gamma_{f}^{2}+\gamma_{p}^{2}\left(1+2e^{\frac{iL}{v_F\hbar}(2\omega-eV_{g,T}-eV_{g,B})}\right)\right],\nonumber\\
&& {\cal G}_{L\downarrow}^r(x,x_2;\omega)==\frac{ie^{-\frac{i}{v_F\hbar}(x-x_2)(\omega-eV_{g,T})}}{\hbar v_F\overline{\Delta}}\left[-1+\gamma_{p}^{2}+\gamma_{f}^{2}\left(1+2e^{\frac{iL}{v_F\hbar}(eV_{g,T}-eV_{g,B})}\right)\right] .
\ea
\end{widetext}
The other spin diagonal components are calculated changing $V_{g,T}\rightarrow V_{g,B}$.

\section{Transmission functions}
\label{tf}
Since the system we are dealing is time-reversal, the transmission functions fulfill $T_{\alpha,\beta}=T_{\beta,\alpha}$. Hereinafter we present the expressions for all transmission functions in distinct cases (single and double QPC), see Ref.~\onlinecite{Dolcini:2011dp}.
\subsection{Single QPC}
In the case where is a single QPC, we have the following transmission functions,  

\ba
&&T_{1(2),3(4)}=\frac{4\gamma_f^2}{(1+\gamma_p^2+\gamma_f^2)^{2}}, \nonumber\\
&&T_{2(1),3(4)}=\frac{(-1+\gamma_p^2+\gamma_f^2)^2}{(1+\gamma_p^2+\gamma_f^2)^{2}},\nonumber\\
&&T_{4(1),3(2)}=\frac{4\gamma_p^2}{(1+\gamma_p^2+\gamma_f^2)^{2}}.
\label{t1}
\ea

\subsection{Two QPC}
\label{td2}

In the case with two QPC, the transmission functions are:   
\ba
&T_{1(2),3(4)}&=\frac{8\gamma_{f}^{2}[-1+\gamma_{f}^2+\gamma_{p}^{2}]^{2}[1+\cos\phi_{MZ}]}{|\overline{\Delta} |^{2}},\nonumber\\
&T_{2(1),3(4)}&=\frac{1}{|\overline{\Delta}|^{2}}\left[(-1+\gamma_{f}^2+\gamma_{p}^{2})^{4}\right.\nonumber\\
&&\left.+16\gamma_f^4-8\gamma_{f}^{2}(-1+\gamma_{f}^2+\gamma_{p}^{2})^{2}\cos\phi_{MZ}\right],\nonumber\\
&T_{4(1),3(2)}&=\frac{8\gamma_{p}^{2}[1+\gamma_{f}^2+\gamma_{p}^{2}]^{2}[1+\cos(\frac{2\omega L}{\hbar v_F}-\phi_{FP})]}{|\overline{\Delta} |^{2}},
\ea

where $\overline{\Delta}=4\gamma_{p}^{2}e^{i\frac{L}{\hbar v_F}(2\omega-eV_{gT}-eV_{gB})}+(1+\gamma_{p}^{2}+\gamma_{f}^{2})^{2}$, and
\ba \label{phases}
\phi_{MZ} &=&\frac{eL(V_{g,B}-V_{g,T})}{v_F\hbar} \nonumber \\
 \phi_{FP} &=&\frac{eL(V_{g,B}+eV_{g,T})}{v_F\hbar}.
 \ea

\end{document}